\documentclass[
 reprint,
 amsmath,amssymb,superscriptaddress,
 aps,prl
]{revtex4-1}
\usepackage{hyperref}

\usepackage{graphicx}
\usepackage{dcolumn}
\usepackage{bm}
\usepackage{dsfont}
\usepackage{natbib}
\usepackage[usenames, dvipsnames]{color}

\newtheorem{lemma}{Lemma}

\DeclareMathOperator*{\argmax}{arg\,max}

\hypersetup{colorlinks=true, citecolor=blue}

\begin{document}

\title{Quantum Delocalised-Interactions}

\author{A.~J.~Paige}\thanks{a.paige16@imperial.ac.uk}
\author{Hyukjoon Kwon}
\author{Selwyn Simsek}
\author{Chris N. Self}
\affiliation{QOLS, Blackett Laboratory, Imperial College London, South Kensington, London, SW7 2AZ, UK.}
\author{Johnnie Gray}
\affiliation{QOLS, Blackett Laboratory, Imperial College London, South Kensington, London, SW7 2AZ, UK.}
\affiliation{Division of Chemistry and Chemical Engineering, California Institute of Technology, Pasadena, California 91125, USA}
\author{M.~S. Kim}
\affiliation{QOLS, Blackett Laboratory, Imperial College London, South Kensington, London, SW7 2AZ, UK.}

\begin{abstract}
Classical mechanics obeys the intuitive logic that a physical event happens at a definite spatial point. Entanglement however, breaks this logic by enabling interactions without a specific location. In this work we study these delocalised-interactions. These are quantum interactions that create less locational information than would be possible classically, as captured by the disturbance induced on some spatial superposition state. We introduce quantum games to capture the effect and demonstrate a direct operational use for quantum concurrence in that it bounds the non-classical performance gain. We also find a connection with quantum teleportation, and demonstrate the games using an IBM quantum processor.
\end{abstract}
\maketitle

Entanglement lies at the heart of the differences between classical and quantum physics. Studying its implications has repeatedly reshaped our understanding of what nature fundamentally allows~\cite{horodecki2009quantum}. In addition to its role in quantum foundations, entanglement is necessary for several types of non-classical advantage~\cite{jozsa2003role,vidal2003efficient,giovannetti2006quantum} and provides the archetypal quantum resource theory~\cite{chitambar2019quantum}. For specific tasks, certain entangled states provide non-classical advantage while others do not. Based on this, entanglement can be divided into different levels of hierarchies, such as steering~\cite{schrodinger1935discussion,wiseman2007steering} and Bell non-locality~\cite{bell1964einstein}. Interestingly, this fundamentally motivated hierarchy has connections to quantum cryptography~\cite{Gisin2002quantum}, with corresponding levels of security for entanglement~\cite{curty2004entanglement}, steering~\cite{branciard2012one}, and Bell non-locality~\cite{mayers1998quantum,acin2007device}.

A key method for studying particular aspects of entanglement is to consider non-local games, where entanglement can provide a non-classical advantage. The archetypal example is the game constructed from the Clauser-Horne-Shimony-Holt (CHSH) test~\cite{clauser1969proposed}. In this CHSH game, Charlie passes two random classical bits $x,y\in\{0,1\}$ to Alice and Bob respectively. Without communicating to each other, Alice and Bob must select and send back bits $a,b\in\{0,1\}$ respectively, and they win the game if $a\oplus b = x\cdot y,$ where $\oplus$ denotes addition modulo $2.$ The best classical strategy gives a win probability of $0.75,$ but using entangled quantum resources they can win with the maximum probability $\frac{1}{2}(1+\frac{1}{\sqrt{2}})\approx0.85.$ Defining and studying games where entanglement provides non-classical performance has been key to improving our understanding of entanglement~\cite{cleve2004consequences,buscemi2012all,branciard2013measurement,regev2015quantum,fritz2012tsirelson,russo2017extended,johnston2016extended,molina2012hedging,dinur2014analytical,cooney2015rank,tavakoli2018semi,khan2018quantum,bennet2012arbitrarily}, since these games neatly encapsulate the often counter-intuitive consequences for information processing governed by the laws of quantum mechanics.

In this work, we study quantum delocalised-interactions, whereby information encoded using non-locally superposed quantum states, is recorded via local interactions whilst causing less disturbance than would be classically possible. This indicates that such interactions cannot be said to happen at a single location. This stands in stark contrast to our classical intuition that interactions happen at unique places, we just might not know where. This non-classical phenomenon has in fact been instrumental in enabling certain quantum protocols~\cite{brodutch2016nonlocal,paige2019quantum}.

In order to characterise delocalised interactions quantitatively, we formulate quantum games and study two particular instances. We establish that the win probabilities of these games are upper bounded in terms of the concurrence for two-qubit states~\cite{hill1997entanglement,wootters1998entanglement}, and the bounds can be saturated for any pure state and a broad class of mixed states. This provides an operational meaning of the concurrence, which has been a widely studied measure of entanglement but is often viewed as a mathematical device. We find that the capacity for non-classical teleportation fidelity~\cite{bennett1993teleporting} guarantees the capacity for non-classical performance in a delocalised-interaction game. We also demonstrate the games using an IBM quantum processor, achieving non-classical performance.

\begin{figure*}
\includegraphics[width=0.78\linewidth]{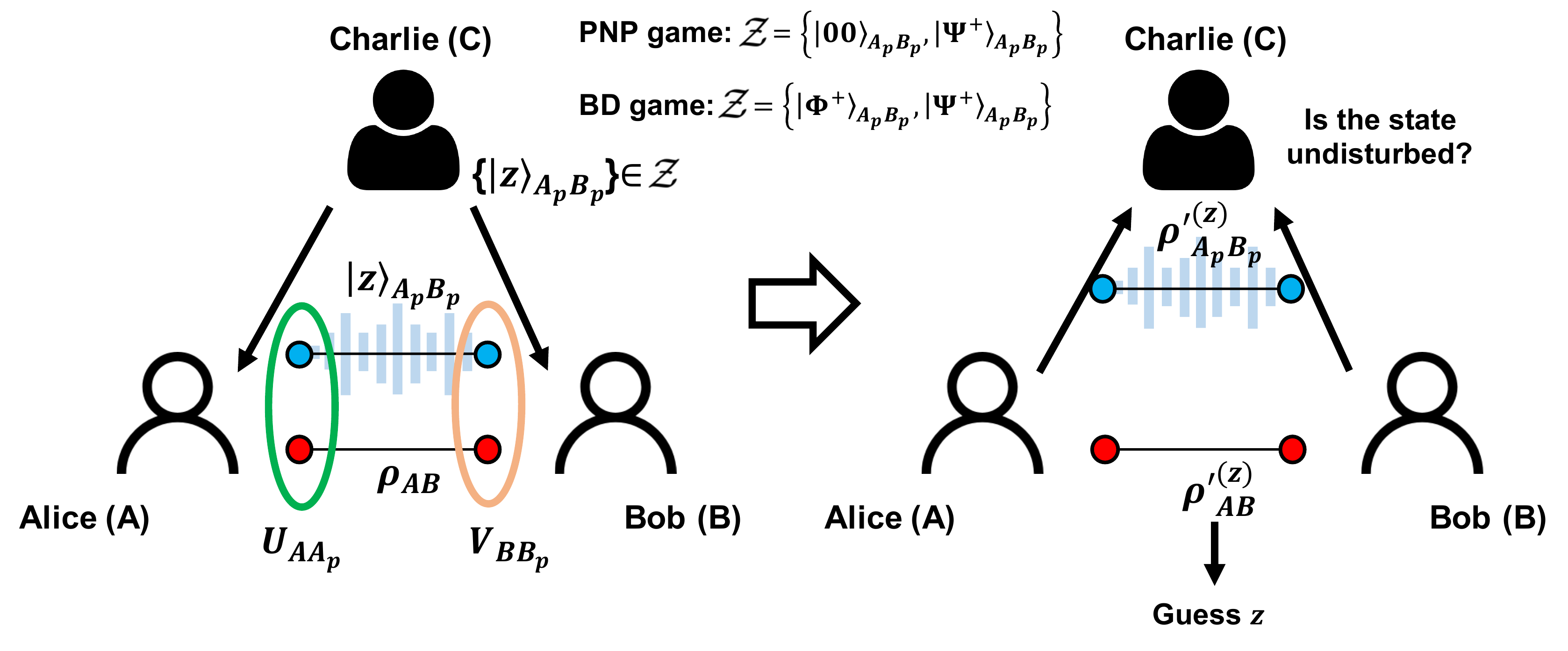}
\caption{Schematic illustration of the quantum delocalised-interaction games as described in the main text, with $\rho_{A_{p}B_{p}}'$ denoting the final state returned to C, and $\rho_{AB}'^{(z)}$ denoting the final state obtained by A and B, which they measure to determine their guess for $z.$ The sets of question states used for the PNP game and the BD game are presented at the top.}
\label{Fig:FormalGame}
\end{figure*}

\emph{Double slit --}
First we illustrate what we mean by delocalised-interactions using the familiar double slit thought experiment. Suppose a game where Charlie (C) either sends a particle through the double slit or does not. Alice (A) standing at one slit together with Bob (B) standing at the other, team up to guess whether C sent the particle or not, without destroying the interference pattern. To win this game, A and B should be able to distinguish between two different states passing through the double-slit, namely a vacuum state $|0\rangle$ and a superposition between spatially separated states $|\psi_{L}\rangle + |\psi_{R}\rangle,$ by locally interacting with the particle. Note these states can also be written as $|00\rangle_{A_{p}B_{p}}$ and $\frac{1}{\sqrt{2}}(|10\rangle_{A_{p}B_{p}} + |10\rangle_{A_{p}B_{p}}),$ where $A_{p}$ and $B_{p}$ are the particle Fock spaces at A and B's locations. If A and B only share classical resources, a perfect record of the existence of the particle is impossible due to the complementarity principle of quantum mechanics. There will be a trade-off, the more information A and B locally record on whether a particle is present, the more they destroy the interference between the different paths by disturbing the superposition state $|\psi_{L}\rangle + |\psi_{R}\rangle$~\cite{englert1996fringe}. On the other hand, if A and B share copies of
a Bell state, for example, $|\Phi^{+}\rangle_{AB}=\frac{1}{\sqrt{2}}(|00\rangle+|11\rangle)_{AB},$ then they can produce a perfect record of when there were particles without affecting the interference pattern. To do this, A and B set up their local
interactions such that the particle flips the local state
as $| 0 \rangle_{A(B)} | 1 \rangle_{A_p(B_p)} \rightarrow | 1 \rangle_{A(B)} | 1 \rangle_{A_p(B_p)}$ and $| 1 \rangle_{A(B)} | 1 \rangle_{A_p(B_p)} \rightarrow | 0 \rangle_{A(B)} | 1 \rangle_{A_p(B_p)},$ while the local states remain the same when the particle is not present. Under this
interaction, the resulting joint state evolves as $|\Phi^{+}\rangle_{AB}(|\psi_{L}\rangle + |\psi_{R}\rangle)\rightarrow\frac{1}{\sqrt{2}}(|01\rangle+|10\rangle)_{AB}(|\psi_{L}\rangle + |\psi_{R}\rangle)=|\Psi^{+}\rangle_{AB}(|\psi_{L}\rangle + |\psi_{R}\rangle)$ when C sent the particle or $|\Phi^{+}\rangle_{AB}|0\rangle\rightarrow|\Phi^{+}\rangle_{AB}|0\rangle$ when C does not send the particle. The interference patterns of the particle have not been disturbed and A and B will have a perfect record of the existence of the particle as their shared outcome states $|\Phi^{+}\rangle_{AB}$ and $|\Psi^{+}\rangle_{AB}$ are perfectly distinguishable.

As illustrated in the double-slit experiment, entanglement allows us to overcome the trade-off between ``information gain via local interaction'' and ``disturbance
in non-local superposition'' i.e., recording information encoded using non-locally superposed quantum states, via local interactions whilst causing less disturbance than would be classically possible. We term this phenomenon delocalised-interactions, as the interaction cannot be known to have definitely happened at either A or B's location, since this would destroy the non-local superposition. We proceed to construct a formal quantum game to quantitatively capture the advantage of sharing entanglement between A and B when demonstrating delocalised-interactions.

\emph{Quantum delocalised-interaction games --} 
We formulate general quantum delocalised-interaction games as follows (illustrated in Fig.~\ref{Fig:FormalGame})
\begin{enumerate}
    \item C prepares a state $|z\rangle_{A_p B_p}$ selected from some finite set of question states $\mathcal{Z}$ with non-zero probability $P_z,$ and sends the subsystems $A_p$ and $B_p$ to A and B, respectively.
    \item A and B attempt to record the information $z$ onto their shared state $\rho_{AB}$ via local controlled unitaries $U_{A A_p}$ and $V_{B B_p},$ then return the subsystems $A_p$ and $B_p$ to $C$.
    \item C checks whether the returned subsystems $A_p B_p$ have been disturbed by performing a projective measurement onto the initial state $|z\rangle_{A_p B_p}.$
    \item A and B perform joint measurements $\Pi_{AB}^{(z_{a})}$ to determine their answer $z_{a}.$
    \item A and B win the game if their answer is correct $z_{a}=z,$ and C's projective measurement returns the initial state $|z\rangle_{A_p B_p}.$
\end{enumerate}
The question states must not be chosen such that A and B cannot distinguish them, and at least one $|z\rangle_{A_{p}B_{p}}\in\mathcal{Z}$ must be entangled. This condition ensures that the games capture the classical trade-off which a quantum delocalised-interaction circumvents.

The probability that A and B win the game is given as
\begin{equation}
    p(\rho_{AB}) = \sum_{z}P_{z}\text{Tr}\big[(\Pi_{AB}^{(z)}\otimes|z\rangle\langle z|)W(\rho_{AB}\otimes|z\rangle\langle z|)W^\dagger\big],
\end{equation}
where $W = U_{A A_p} \otimes V_{B B_p}.$ We shall use the superscript form $p^{\text{m}}$ to denote the maximum of this quantity over all choices of measurements $\Pi_{AB}$ and controlled unitaries $U_{A A_p},V_{B B_p},$ and we shall use subscripts to distinguish specific instances.

\emph{Particle/No-Particle game --} 
The double-slit scenario can now be simplified into an example of a quantum delocalised-interaction game. In this case, $\mathcal{Z} = \{|\text{p}\rangle, |\text{np}\rangle\}$ with $P_{\text{p}} = 1/2 = P_{\text{np}}$ and we take $|\text{p}\rangle =\frac{1}{\sqrt{2}}(|01\rangle_{A_{p}B_{p}}+|10\rangle_{A_{p}B_{p}}),$ and $|\text{np}\rangle =|00\rangle_{A_{p}B_{p}},$ which represent the states after passing the double-slit depending on whether C sends (p) or does not send (np) the particle.

We also choose to work with the interaction only happening if the particle exists in the local subsystem, since a unitary in the absence of a particle physically corresponds to free evolution which we can simply factor out. Hence $U_{A A_p} = \mathds{1}_A \otimes|0\rangle_{A_p}\langle0| + U_{A} \otimes |1\rangle_{A_{p}}\langle1|$
and $V_{B B_p} = \mathds{1}_B \otimes|0\rangle_{B_p}\langle0| + V_{B} \otimes |1\rangle_{B_{p}}\langle1|.$ The overall interaction then can be written as
\begin{equation}
\begin{split}
    W&=\mathds{1}_{AB}\otimes|00\rangle_{A_{p}B_{p}}\langle00| + U_{A}\otimes\mathds{1}_{B}\otimes|10\rangle_{A_{p}B_{p}}\langle10| \\
    & +\mathds{1}_{A}\otimes V_{B}\otimes|01\rangle_{A_{p}B_{p}}\langle01| + U_{A}\otimes V_{B}\otimes|11\rangle_{A_{p}B_{p}}\langle11| .
\end{split}
\end{equation}

We refer to this game as the Particle/No-Particle (PNP) game and we find that the maximum obtainable win probability for a pure two-qubit state is given as
\begin{equation}\label{Eq:GNPPureState}
    p_{\text{pnp}}^{\text{m}}(|\psi\rangle_{AB})= \frac{3}{4}+\frac{1}{4}C(|\psi\rangle_{AB}),
\end{equation}
where $C(|\psi\rangle_{AB})=2\sqrt{\lambda_0\lambda_1}$ with $\lambda_{i}$ denoting the Schmidt coefficients, is the well-known concurrence entanglement monotone~\cite{hill1997entanglement,wootters1998entanglement} (proof in Ref.~\cite{Supplemental}), which is zero for all separable states, giving the classical bound as $\frac{3}{4}.$ We can therefore view the game as providing a direct operational meaning of pure state concurrence.

This result has interesting implications, for instance one might have thought that A and B would be helped by allowing a pre-processing step where they have temporary access to all the states they will use, and can apply entanglement distillation. However, using the concurrence result we can show that this would not increase their win probability. Consider A and B granted pre-processing access to $N$ copies of the qubit state $|\psi\rangle,$ from which they distil $m$ copies of the maximally entangled state and $N-m$ pure separable states. Then when the game starts they use these new states one by one, and win $m$ cases with probability $1$ and $N-m$ cases with the maximum classical win probability $\frac{3}{4}.$ It is known that in the asymptotic limit of large $N$ we have $m=NE(|\psi\rangle),$ where $E(|\psi\rangle)$ is the entanglement entropy~\cite{bennett1996concentrating}. This means the win probability for the outlined distillation strategy will be bounded by $\frac{1}{N}[NE(|\psi\rangle) + \frac{3}{4}(N-NE(|\psi\rangle)]=\frac{3}{4}+\frac{1}{4}E(|\psi\rangle).$ However by using the original states they would obtain $\frac{3}{4}+\frac{1}{4}C(|\psi\rangle),$ and it is known that $C(|\psi\rangle)\geq E(|\psi\rangle).$ Therefore the distillation does not provide improvement.

To generalise Eq.~(\ref{Eq:GNPPureState}) to mixed states we use the fact that the maximum win probability is a convex function $p^{\text{m}}(\sum_{i}r_{i}\rho^{(i)})\leq \sum_{i}r_{i} p^{\text{m}}(\rho^{(i)}),$ which can be intuitively understood as follows. Consider A and B being either given copies of a known state $\sum_{i}r_{i}\rho^{(i)},$ or given labelled copies of known states $\rho^{(i)}$ where the number of each is in proportion to $r_{i}.$ From the second case they can reproduce the first case by simply ignoring the labels, therefore in the second case they must be able to obtain at least as high a win probability as in the first case, hence the convexity result. Using this, combined with the fact that $C(\rho_{AB}) = \inf \sum_{i} q_i C(|\psi_i\rangle_{AB}),$ we can extend Eq.~(\ref{Eq:GNPPureState}) to a bound for mixed states, giving
\begin{equation}\label{Eq:GNPMixedBound}
    p_{\text{pnp}}^{\text{m}}(\rho_{AB}) \leq \frac{3}{4} + \frac{C(\rho_{AB})}{4}.
\end{equation}
Since the concurrence has an analytic closed form, we can now easily calculate a bound on the win probability gain for any two-qubit state.

From this we can also view the game as providing a direct operational meaning of concurrence for mixed states that saturate the bound. It is therefore natural to ask whether the bound can be tight for mixed states. The answer is yes, as we found that it saturates for mixtures of two Bell states~\cite{Supplemental}. However, this is not true for all mixed states. An informative example is given by Werner-like states~\cite{werner1989quantum} $\rho_{AB}=a|\psi^{k}\rangle_{AB}\langle\psi^{k}|+\frac{1-a}{4}\mathds{1}_{AB},$ where $0\leq a\leq 1$ and $|\psi^{k}\rangle$ is chosen as one of the four Bell states. We shall show that this state does not saturate the concurrence bound.

To understand and prove this behaviour we note that the mixedness of a state can degrade its record quality. Consider the extreme example of the maximally mixed state $\mathds{1}_{AB}/n$. It is clear that if A and B try to unitarily encode the presence of a particle in this state then they will not gain information. This inability of the state to acquire information is what we intuitively mean when we say it has bad record quality. We capture the general effect via the bound
\begin{equation}
    p_{\text{pnp}}^{\text{m}}(\rho_{AB}) \leq \frac{1}{2}+\frac{1}{2}T_{c}(\lambda^{\uparrow},\lambda^{\downarrow}),
\end{equation}
where we denote the classical trace distance $T_{c}(p,q)=\frac{1}{2}\sum_{i}|p_{i}-q_{i}|$ for probability vectors $p,q$ defined over the same index set, and $\lambda^{\uparrow}$ is the vector of eigenvalues of $\rho$ arranged in ascending order and including any zero values. For the $\mathds{1}_{AB}/n$ example, we see that the win probability cannot exceed $\frac{1}{2},$ i.e., the best they can do is just guess. The proof of the bound proceeds via the lemma $T(\rho,\sigma) \leq T_{c}(\lambda^{\uparrow},\mu^{\downarrow}),$ where $\mu^{\downarrow}$ is the vector of eigenvalues of $\sigma$ in descending order (see Ref.~\cite{Supplemental} for details).

Returning to the Werner-like states, we find that this record quality bound can be saturated. This can be demonstrated with $U_A=X_A=|0\rangle_{A}\langle1|+|1\rangle_{A}\langle0|,$ and $V_B=\pm X_B,$ where the sign is chosen to match the sign of $\langle\psi^{k}|X_{A}X_{B}|\psi^{k}\rangle.$ This gives $p_{\text{pnp}}=\frac{1}{2}(1+a),$ which exactly saturates the record bound and is therefore an optimal tactic. This record bound is below the concurrence bound for all $a<1,$ and therefore the Werner-like states cannot in general saturate the concurrence bound.

Since the Werner-like state is entangled for $a>\frac{1}{3},$ these results indicate that entanglement is not sufficient to observe nonclassical advantage in the PNP quantum game. Additionally we note that the capacity for Bell non-locality is not necessary for a state to demonstrate non-classical performance, since there is a local model for projective measurements for $a\lesssim0.66$ \cite{acin2006grothendieck}. We note that this appears to hold even if we allow A and B to use additional pure classical states (see \cite{Supplemental} for results of numerics using qutip~\cite{johansson2012qutip,johansson2013qutip}).

\emph{Bell distinguishing game --} We now study a modified game that indicates an even stronger connection with concurrence. In the PNP game considered above, the no particle state $|\text{np}\rangle =|00\rangle_{A_{p}B_{p}},$ has no spatial superposition which can be damaged by the local measurements. To move away from this, we can consider replacing $|00\rangle_{A_{p}B_{p}},$ with the Bell state $|\Phi^{+}\rangle_{A_{p}B_{p}}=\frac{1}{\sqrt{2}}(|00\rangle_{A_{p}B_{p}}+|11\rangle_{A_{p}B_{p}}).$ So Alice and Bob are now tasked with distinguishing two Bell states $|\Psi^{+}\rangle$ and $|\Phi^{+}\rangle$ whilst trying to return them undamaged. We shall refer to this as the Bell-Distinguishing (BD) game. It is noteworthy that this task can be viewed as detecting local bit-flip errors, where in contrast to a conventional syndrome measurement~\cite{roffe2019quantum} one is using two ancilla modes, each of which can only interact with its local part of the system.

For two-qubit states we again find that the concurrence quantifies the maximum obtainable win probability~\cite{Supplemental}, via
\begin{equation}
    p_{\text{bd}}^{\text{m}}(|\psi\rangle_{AB})= \frac{1}{2} + \frac{1}{2}C(|\psi\rangle_{AB}),
\end{equation}
and thus we have the general bound
\begin{equation}\label{Eq:BDConBound}
    p^{\text{m}}_{\text{bd}}(\rho_{AB})\leq \frac{1}{2} + \frac{1}{2}C(\rho_{AB}).
\end{equation}

Unlike for the PNP game, Werner-like states can saturate the concurrence bound, and in fact we find that the well studied Bell diagonal states~\cite{lang2010quantum,quan2016steering,kay2012using,kent1999optimal,cen2002local,ren2014correlation} can all saturate the bound. In order to prove this, we note that there exists a tactic with win probability at least equal to the fully entangled fraction (singlet fraction) $\mathcal{F}(\rho)=\max_{\psi}\langle\psi|\rho|\psi\rangle,$ where the maximum is taken over all maximally entangled states of the system. A and B can achieve this by adopting the optimal tactic for the maximally entangled pure state $|\psi^*\rangle=\argmax_{\psi}\langle\psi|\rho|\psi\rangle.$ Since all entangled Bell diagonal states have concurrence $C(\rho)=2\mathcal{F}(\rho)-1,$ so the tactic outlined above leads to $p_{\text{bd}}(\rho)=\frac{1}{2}+\frac{1}{2}C(\rho)$ thus saturating the bound.

The above outlined tactic also produces an interesting corollary regarding quantum teleportation~\cite{bennett1993teleporting}, namely that all entangled two-qubit states capable of non-classical teleportation fidelity are also capable of non-classical performance in the BD game, since it is known that a two-qubit state can achieve non-classical teleportation fidelity if and only if $\mathcal{F}>\frac{1}{2}$~\cite{horodecki1999general,badziag2000local}. It would be an interesting open question to study whether the converse
statement is true. We conjecture that this might be the
case by numerically verifying that examples of entangled two-qubit states with $\mathcal{F}<\frac{1}{2}$~\cite{horodecki2001mixed} do not show non-classical BD performance.

\begin{figure}
\includegraphics[width=\linewidth]{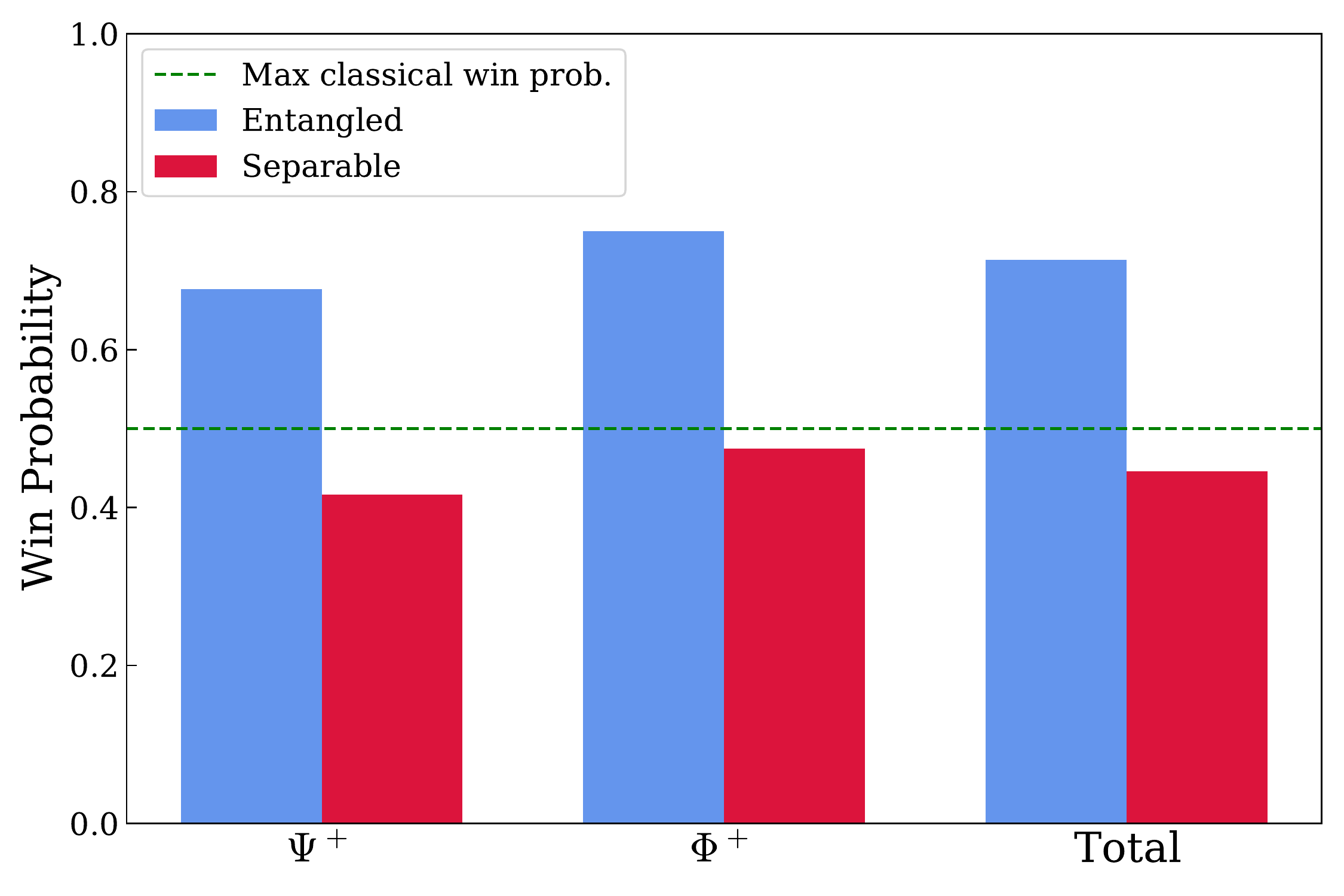}
\caption{Plot of results for the BD game, calculated from \emph{Paris} device measurements. The total win probability is calculated for equal probability of sending either state. The blue bars are for an entangled initial resource state, the red for separable, and the green line is the maximum classical win probability.}
\label{fig:BDParis}
\end{figure}

\emph{IBM machine demonstration--}  We tested the delocalised-interaction games using the IBM superconducting quantum processor \emph{Paris}. We implemented both the BD and PNP games by initially sharing a Bell state between A and B and by employing controlled bit flips as local interaction unitary operators (for details see Ref.~\cite{Supplemental}). A key simplifying aspect of this approach is that A and B do not have to perform a joint measurement at the end to determine their answer. The two possible states are $\frac{1}{\sqrt{2}}(|00\rangle_{AB}+|11\rangle_{AB})$ and $\frac{1}{\sqrt{2}}(|01\rangle_{AB}+|10\rangle_{AB}),$ so they can simply measure in the local $Z$ basis and base their answer on the joint parity of their results. Therefore they only require LOCC rather than joint measurements.

The results for the BD game are illustrated in Fig.~\ref{fig:BDParis}, where alongside the results for the entangled initial state, we include results for the separable initial state $|00\rangle$ for comparison. The entangled win probability achieved was $0.71,$ which is far from the ideal but violates the classical limit of $0.5.$ This demonstrates a usable concurrence of $0.42$ and thus a convincing delocalised-interaction. For the standard PNP game we could not demonstrate non-classical performance, but altering the game by increasing the probability a particle is sent $P_{\text{p}},$ we were able to establish non-classical performance, although we cannot currently relate this violation directly to an entanglement measure (details in Ref.~\cite{Supplemental}).

Note that this is not an ideal demonstration of delocalised-interactions. Imperfections in the device's behaviour could in principle be used to account for the non-classical result. Delocalised-interactions are subject to the usual loopholes that plague demonstrations of non-local effects~\cite{larsson2014loopholes}. Potentially these could be addressed by future experiments with photonic qubits~\cite{kok2007linear}.

\emph{Conclusions--}
In this work we studied the concept of delocalised-interactions. Information encoded using non-locally superposed quantum states, is recorded via local interactions whilst disturbing the superposition less than would be classically possible. This phenomenon has interesting foundational implications regarding events not requiring unique locations and has also been a key component for certain quantum protocols~\cite{brodutch2016nonlocal,paige2019quantum}. In order to systematically study this quantum effect, we introduced and investigated quantum games for which non-classical performance demonstrates delocalised-interactions. This enabled us to prove a direct operational use of concurrence in bounding the non-classical win probabilities, and a connection with quantum teleportation. Our work can spur further research building from the tools and ideas introduced here, such as generalising to higher dimensions or multipartite settings, and establishing the exact nature of the connection with quantum teleportation. Finally, the delocalised-interaction games were demonstrated on an IBM superconducting quantum processor, finding non-classical performance.

\begin{acknowledgements}
\emph{Acknowledgments ---}
We acknowledge insightful discussions with Benjamin Yadin, David Jennings, Adam Callison, and Thomas Hebdige. AP and SS are funded by the EPSRC Centre for Doctoral Training in Controlled Quantum Dynamics. We thank the Royal Society, the KIST Open Lab programme and the Samsung GRP grant for financial support.
\end{acknowledgements}

\bibliographystyle{apsrev4-1}

\onecolumngrid
\newpage

\section{Supplemental Material}

\subsection{PNP game for pure states}\label{appen:PNPforPure}
To derive the entanglement bound we start by finding the maximum win probability for a pure state. Maximising over the choice of measurements the win probability is
\begin{equation}\label{eq:PureStateWinProb}
    p_{\text{pnp}}(|\psi\rangle_{AB}\langle\psi|) =\frac{1}{2}[1 + \text{eigs}_{+}(M)],
\end{equation}
where $M = \frac{1}{4}(|\psi_u\rangle+|\psi_v\rangle)(\langle\psi_u|+\langle\psi_v|)-|\psi\rangle\langle\psi|,$ and we are writing $U_A|\psi\rangle=|\psi_u\rangle,$ $V_B|\psi\rangle=|\psi_v\rangle.$

We now need to maximise over the choice of unitaries. Note that if $|\psi_u\rangle + |\psi_v\rangle$ is parallel with $|\psi\rangle,$ then it follows that $M\leq0,$ and so $p_{\text{pnp}}\leq\frac{1}{2}.$ This means we just need to consider the case where this is not true, such that $M$ has rank $2.$ The two non-zero eigenvalues must sum to $\text{Tr}(M),$ so we write $m_1+m_2=\frac{1}{2}(1+\text{Re}\langle\psi_u|\psi_v\rangle)-1,$ from which we have $-1 \leq m_1+m_2 \leq 0.$ This means that we can have at most one positive eigenvalue, and the maximization will choose unitaries that maximise the magnitude of this positive eigenvalue.

We now consider an operator of the form $M=K_1|\psi\rangle\langle\psi|K_1^\dagger - K_2|\psi\rangle\langle\psi|K_2^\dagger,$ (we are performing a more general treatment as the result shall prove useful later as well). For this we derive an expression for its eigenvalues. We write
\begin{equation*}
\begin{split}
    m (\alpha K_1|\psi\rangle + \beta K_2 |\psi\rangle) = (K_1|\psi\rangle\langle\psi|K_1^\dagger - K_2|\psi\rangle\langle\psi|K_2^\dagger)(\alpha K_1|\psi\rangle + \beta K_2 |\psi\rangle) \\
    = (\alpha\langle\psi|K_1^\dagger K_1|\psi\rangle + \beta \langle\psi|K_1^\dagger K_2 |\psi\rangle)K_1|\psi\rangle - (\alpha\langle\psi|K_2^\dagger K_1|\psi\rangle + \beta \langle\psi|K_2^\dagger K_2 |\psi\rangle)K_2|\psi\rangle.
\end{split}
\end{equation*}

Under the assumption that $K_1|\psi\rangle$ and $K_2|\psi\rangle$ are not proportional, and defining $k_{ij}=\langle\psi|K_i^\dagger K_j|\psi\rangle,$ we obtain the following two equations
\begin{equation*}
    m\alpha = (\alpha k_{11} + \beta k_{12}),
\end{equation*}
\begin{equation*}
    m\beta = -(\alpha k_{21} + \beta k_{22}).
\end{equation*}
Combining these to eliminate $\alpha$ and $\beta$ we find
\begin{equation*}
    m^2+(k_{22}-k_{11})m-(k_{11}k_{22}-k_{12}k_{21}) = 0.
\end{equation*}
The solutions to this quadratic equation are then found to be
\begin{equation}\label{Eq:generalEigenVal}
    m = \frac{1}{2}\big(k_{11}-k_{22}\pm\sqrt{(k_{22}+k_{11})^2-4k_{12}k_{21}}\big).
\end{equation}

Applying this result to the case at hand we find that the largest eigenvalue is given by
\begin{equation*}
    m(k,\kappa) = \frac{(k-1)+\sqrt{(k+1)^2-4\kappa}}{2},
\end{equation*}
where $k=\langle\psi|K^\dagger K|\psi\rangle,$ and $\kappa = |\langle\psi|K|\psi\rangle|^{2},$ and $K=(U_{A}+V_{B})/2.$ We write these out explicitly as
\begin{equation*}
    k= \frac{1}{2}(1+\text{Re}\langle\psi|U_{A}^\dagger V_{B}|\psi\rangle),
\end{equation*}
\begin{equation*}
    \kappa=\frac{1}{4}(u^2+v^2+2uv\cos\Delta\phi),
\end{equation*}
where we are using $\langle\psi|U_{A}|\psi\rangle = e^{i\phi_{A}}u$ and $\langle\psi|V_{B}|\psi\rangle = e^{i\phi_{B}}v$ with $u,v\in\mathbb{R}_{\geq0},$ and $\Delta\phi\equiv\phi_A-\phi_B.$

We now insert a resolution of the identity $\mathds{1}=|\psi\rangle\langle\psi|+P_{\perp}$ to rewrite $k$ as 
\begin{equation*}
    k = \frac{1}{2}(1+uv\cos(\Delta\phi) +\text{Re}\langle\psi|U_{A}^\dagger P_{\perp}V_{B}|\psi\rangle ).
\end{equation*}
From this we write
\begin{equation*}
    k  \leq \frac{1}{2}[1+ uv\cos\Delta\phi + G(|\psi\rangle)] \equiv k_{b},
\end{equation*}
where
\begin{equation*}
    G(|\psi\rangle)\equiv\max_{U_{A},V_{B}}|\langle\psi|U_{A}^\dagger P_{\perp}V_{B}|\psi\rangle|,
\end{equation*}
is a state dependent constant.

We shall only consider values of $G(|\psi\rangle)<1$ since the maximum win probability is $1$ for $G=1.$ Now since increasing $k$ can only increase $m$ we can write $m(k,\kappa) \leq m(k_b,\kappa).$ This is important as by construction $m(k_b,\kappa),$ only depends on the three variables $u,v,\Delta\phi,$ and this enables us to maximise via taking partial derivatives.

We start by writing
\begin{equation*}
    \tilde{m}(k,\kappa)\equiv m(k_b,\kappa) = \frac{(k_b-1)+\sqrt{(k_b+1)^2-4\kappa}}{2},
\end{equation*}
where we have $k_{b} = \frac{1}{2}[1+ uv\cos\Delta\phi + G(|\psi\rangle)],$ and $\kappa=\frac{1}{4}(u^2+v^2+2uv\cos\Delta\phi).$
    
Taking partial derivatives w.r.t. $\Delta\phi,u,v$ via the chain rule we find
\begin{equation*}
    \frac{\partial \tilde{m}}{\partial\Delta\phi} =-\frac{1}{4}uv\sin\Delta\phi (1+\frac{k_b-1}{\sqrt{(k_b+1)^2-4\kappa}}),
\end{equation*}
\begin{equation*}
    \frac{\partial \tilde{m}}{\partial u} = \frac{1}{4}(1+\frac{k_b-1}{\sqrt{(k_b+1)^2-4\kappa}})(v\cos\Delta\phi ) - \frac{1}{2\sqrt{(k_b+1)^2-4\kappa}}u,
\end{equation*}
\begin{equation*}
    \frac{\partial \tilde{m}}{\partial v} = \frac{1}{4}(1+\frac{k_b-1}{\sqrt{(k_b+1)^2-4\kappa}})(u\cos\Delta\phi ) - \frac{1}{2\sqrt{(k_b+1)^2-4\kappa}}v.
\end{equation*}
Setting these expressions equal to zero we obtain the three equations
\begin{equation}\label{Eq:pd1}
    0 = X uv \sin\Delta\phi ,
\end{equation}
\begin{equation}\label{Eq:pd2}
    0 = Xv\cos\Delta\phi - Yu,
\end{equation}
\begin{equation}\label{Eq:pd3}
    0 = Xu\cos\Delta\phi - Yv,
\end{equation}
where $X\equiv\frac{1}{2}(1+\frac{k_b-1}{\sqrt{(k_b+1)^2-4\kappa}}),$ and $Y\equiv\frac{1}{\sqrt{(k_b+1)^2-4\kappa}}.$

First we observe that $X$ cannot equal zero. To see this we set it to zero and solve by squaring it to arrive at $k_{b}=\kappa,$ but this does not make the original term zero. We also note that $Y\equiv\frac{1}{\sqrt{(k_b+1)^2-4\kappa}}$ cannot equal zero either.

It follows from this that to satisfy Eq.~(\ref{Eq:pd1}) we must take $u=0,$ $v=0,$ or $\sin\Delta\phi=0.$ If we consider $u=0$ then Eq.~(\ref{Eq:pd3}) implies $v=0.$ Similarly if we consider $v=0,$ then Eq.~(\ref{Eq:pd2}) implies $u=0.$ So in both cases we have $u=v=0,$ and this naturally makes the choice of $\Delta\phi$ irrelevant. From this we see that the only choice we now have to consider is $\sin\Delta\phi=0.$

Taking this case we write Eq.~(\ref{Eq:pd2}) and Eq.~(\ref{Eq:pd3}) as $\pm Xv=Yu$ and $\pm Xu=Yv$ respectively. Now if either $\pm Xv=Yu=0$ or $\pm Xu=Yv=0,$ then we quickly see this implies $u=v=0,$ so we just need to consider the option of $\pm Xv=Yu\neq0$ and $\pm Xu=Yv\neq0.$ In this case we can divide through and get $u^2=v^2$ so $u=\pm v.$ Substituting back we have $(X\pm Y)u=0$ but we find that $(X\pm Y)\neq0,$ so we again arrive back at $u=0,$ $v=0,$ which means $\pm Xv=Yu=0$ and $\pm Xu=Yv=0,$ so we have a contradiction.

Putting all this together we have found that the only turning point solution is given by $u=v=0.$ This gives $\kappa=0$ and $k_b=\frac{1}{2}(1+G(|\psi\rangle)).$ And plugging this into the original equation we arrive at the result
\begin{equation*}
    \tilde{m}^{\max} = \frac{1}{2}(1+G(|\psi\rangle)).
\end{equation*}

To verify that this is a maximum we first note that because we are interested in the line $u=v=0$ for all $\Delta\phi$ then we simply have a 2D problem in each plane defined by a fixed value of $\Delta\phi.$ Thus we apply the two dimensional second partial derivative test.

The required second-order partial derivatives are found to be
\begin{equation*}
    \frac{\partial^2 \tilde{m}(u=v=0)}{\partial u^2}= -\frac{1}{2+G},
\end{equation*}
\begin{equation*}
    \frac{\partial^2 \tilde{m}(u=v=0)}{\partial v^2}= -\frac{1}{2+G},
\end{equation*}
\begin{equation*}
    \frac{\partial^2 \tilde{m}(u=v=0)}{\partial u\partial v}=\frac{\partial^2 \tilde{m}(u=v=0)}{\partial v\partial u}= \big(\frac{1}{4}+\frac{G-1}{G+3}\big)\cos\Delta\phi.
\end{equation*}
From which one finds the determinant of the Hessian matrix as
\begin{equation*}
    D(u,v) = \frac{1}{(2+G)^2}-(\frac{1}{4}+\frac{G-1}{G+3}\big)^2 \cos^2\Delta\phi.
\end{equation*}
Now since $D(u,v)>0$ and $\frac{\partial^2 \tilde{m}(u=v=0)}{\partial u^2}<0$ for all valid values of $\Delta\phi$ and $G,$ the second partial derivative test informs us that we have found a maximum.

When we do this we find
\begin{equation*}
    m(k,\kappa) \leq m(k_b,\kappa) \leq \frac{1}{2}[1+G(|\psi\rangle)].
\end{equation*}
Using this with Eq.~(\ref{eq:PureStateWinProb}) we arrive at
\begin{equation*}
    p_{\text{pnp}}(|\psi\rangle_{AB}) \leq\frac{3}{4} + \frac{G(|\psi\rangle_{AB})}{4}.
\end{equation*}

This bound is obtainable for all pure states. In order to see this we note that if we only consider unitaries that map the initial state to an orthogonal state such that $\langle\psi|U_{A}|\psi\rangle=0$ and $\langle\psi|V_{B}|\psi\rangle=0.$ Then the win probability becomes $\frac{3}{4}+\frac{1}{4}\text{Re}[\text{Tr}(U_{A}|\psi\rangle_{AB}\langle\psi|V_{B}^\dagger)],$ which by inserting the previous resolution of identity becomes $\frac{3}{4}+\frac{1}{4}\text{Re}[\langle\psi|U_{A}P_{\perp}V_{B}^\dagger|\psi\rangle].$ We can always choose phases such that $\langle\psi|U_{A}P_{\perp}V_{B}^\dagger|\psi\rangle$ is real, and hence we see that all we need to do is find the best choice of $U_A,V_B$ subject to the orthogonality constraint, and we thus obtain the optimal win probability
\begin{equation*}
    p_{\text{pnp}}^{\text{m}}(|\psi\rangle_{AB})= \frac{3}{4}+\frac{1}{4}G(|\psi\rangle_{AB}),
\end{equation*}
where we use the superscript $\text{m}$ to denote the maximum obtainable value.

Now we prove that for qubit states $G(\rho_{AB})=C(\rho_{AB}).$ We start by defining the orthogonal basis states
\begin{equation*}
\begin{split}
    |\psi\rangle &= \sqrt{r} |00\rangle + \sqrt{1-r} |11\rangle,
\\
    |\psi_1\rangle &= \sqrt{r} |10\rangle + \sqrt{1-r} |01\rangle, 
\\
    |\psi_2\rangle &= \sqrt{1-r} |10\rangle - \sqrt{r} |01\rangle.
\\
    |\psi_3\rangle &= \sqrt{1-r} |00\rangle - \sqrt{r} |11\rangle, 
\end{split}
\end{equation*}
We can then write $P_{\perp}=\sum_i|\psi_i\rangle\langle\psi_i|.$

We parametrize the unitaries in terms of the identity and Pauli operators as $U_{A}=e^{i\phi_A}\mathbf{a}\cdot\boldsymbol{\sigma}_{A},$ where $\boldsymbol{\sigma}=(\mathds{1},X,Y,Z)^{T},$ $\mathbf{a}=(a_0,ia_1,ia_2,ia_3)^{T},$ $a_i\in\mathbb{R}$ and $\sum_i a_i^2 = 1.$ Similarly we write $V_{B}=e^{i\phi_B}\mathbf{b}\cdot\boldsymbol{\sigma}_{B}.$ We also define the three vectors $\vec{a}=(a_1,a_2,a_3)^T,$ $\vec{b}=(b_1,b_2,b_3)^T,$ and $\vec{\sigma}=(X,Y,Z)^T,$

We wish to evaluate
\begin{equation}\label{eq:expandingG}
\begin{split}
    \langle\psi|U_{A}^\dagger P_{\perp} V_{B}|\psi\rangle & = e^{i(\phi_B-\phi_A)} \langle\psi|\vec{a}\cdot\vec{\sigma} P_{\perp} \vec{b}\cdot\vec{\sigma}|\psi\rangle, \\
    & =e^{i(\phi_B-\phi_A)}( \langle\psi|\vec{a}\cdot\vec{\sigma}|\psi_{1}\rangle\langle\psi_{1}|\vec{b}\cdot\vec{\sigma}|\psi\rangle + \langle\psi|\vec{a}\cdot\vec{\sigma}|\psi_{2}\rangle\langle\psi_{2}|\vec{b}\cdot\vec{\sigma}|\psi\rangle + \langle\psi|\vec{a}\cdot\vec{\sigma}|\psi_{3}\rangle\langle\psi_{3}|\vec{b}\cdot\vec{\sigma}|\psi\rangle).
\end{split}
\end{equation}
To do so we calculate
\begin{equation*}
\begin{split}
    \langle\psi_{1}|\vec{a}\cdot\vec{\sigma}|\psi\rangle &= a_1 + i a_2 (2r - 1)),
\\
    \langle\psi_{2}|\vec{a}\cdot\vec{\sigma}|\psi\rangle &= 2 i a_2 \sqrt{r(1-r)},
\\
    \langle\psi_{3}|\vec{a}\cdot\vec{\sigma}|\psi\rangle &= 2 a_3 \sqrt{r(1-r)},
\\
    \langle\psi_{1}|\vec{b}\cdot\vec{\sigma}|\psi\rangle &= 2 b_1 \sqrt{r(1-r)},
\\
    \langle\psi_{2}|\vec{b}\cdot\vec{\sigma}|\psi\rangle &= -i b_2 + b_1(1-2r),
\\
    \langle\psi_{3}|\vec{b}\cdot\vec{\sigma}|\psi\rangle &= 2 b_3 \sqrt{r(1-r)}.
\end{split}
\end{equation*}

We use these to evaluate Eq.~(\ref{eq:expandingG}) and find
\begin{equation*}
\begin{split}
    \langle\psi|U_{A}^\dagger P_{\perp} V_{B}|\psi\rangle & = e^{i(\phi_B-\phi_A)}\big[ 2 b_1 \sqrt{r(1-r)}(a_1 - i a_2 (2r - 1)) \\ 
    & + (-i b_2 + b_1(1-2r)) (-2 i a_2 \sqrt{r(1-r)}) + 4 a_3 b_3 r (1-r) \big], \\
    & =2\sqrt{r(1-r)}e^{i(\phi_B-\phi_A)}( a_1b_1 - a_2b_2 + 2 a_3 b_3 \sqrt{r(1-r)} ).
\end{split}
\end{equation*}
Taking the modulus of this gives
\begin{equation*}
    |\langle\psi|V_{B}^\dagger P_{\perp} U_{A}|\psi\rangle|
    = |2\sqrt{r(1-r)}(a_1b_1 - a_2b_2 + 2 a_3 b_3 \sqrt{r(1-r)} )|.
\end{equation*}
By writing $\vec{v}_1=(a_1,a_2,a_3)^T,$ $\vec{v}_2=(b_1,-b_2,2\sqrt{r(1-r)} b_3)^T,$ we can rewrite the right-hand side as $2\sqrt{r(1-r)}|\vec{v}_1\cdot\vec{v}_2|.$ The Cauchy-Schwarz inequality gives $|\vec{v}_1\cdot\vec{v}_2|\leq|\vec{v}_1||\vec{v}_2|,$ and it is straightforward to show $|\vec{v}_1|\leq 1$ and $|\vec{v}_2|\leq 1,$ which leads us to $|\langle\psi|V_{B}^\dagger P_{\perp} U_{A}|\psi\rangle| \leq 2\sqrt{r(1-r)}.$ Since this maximum is clearly obtainable by for instance setting $a_1=b_1=1$ and the other terms to zero we conclude
\begin{equation*}
    G(|\psi\rangle_{AB})=\max_{U_{A},V_{B}}|\langle\psi|U_{A}^\dagger P_{\perp}V_{B}|\psi\rangle| = 2\sqrt{r(1-r)} = C(|\psi\rangle_{AB}),
\end{equation*}
where we have identified $2\sqrt{r(1-r)}$ as the pure state concurrence~\cite{hill1997entanglement,wootters1998entanglement}, and this also coincides with the negativity~\cite{vidal2002computable}. Now since the concurrence can be defined via a convex roof extension, we have that
\begin{equation*}
    G(\rho_{AB})=C(\rho_{AB}).
\end{equation*}

\subsection{PNP record quality bound}\label{appen:RecordBound}
First we note that one can rewrite the win probability in terms of a trace distance but care is needed on account of $\tilde{\sigma}_{AB}$ not being normalised. To see this we perform the standard separation of positive and negative eigenvalues by writing $\tilde{\sigma}-\rho=Q-S,$ so that $|\tilde{\sigma}-\rho|=Q+S,$ and therefore $T(\tilde{\sigma},\rho)=\frac{1}{2}\text{Tr}|\tilde{\sigma}-\rho|= \frac{1}{2}(\text{Tr}Q+\text{Tr}S).$ We now use $\text{Tr}(\tilde{\sigma}-\rho)=\text{Tr}\tilde{\sigma}-1=\text{Tr}Q-\text{Tr}S,$ to find that $\text{Tr}Q = T(\tilde{\sigma},\rho)-\frac{1}{2}+\frac{1}{2}\text{Tr}\tilde{\sigma}.$ Finally the fact that $\text{Tr}Q=\text{eigs}_{+}(\tilde{\sigma}-\rho),$ leads through to
\begin{equation}\label{Eq:MaxWinProbAltForm}
    p_{\text{pnp}}(\rho_{AB})=\frac{1}{4}+\frac{1}{2}T(\tilde{\sigma}_{AB},\rho_{AB})+ \frac{1}{4}\text{Tr}[\tilde{\sigma}_{AB}].
\end{equation}

With this we now write $\tilde{\sigma}_{AB}=\sigma_{AB}-\tilde{\sigma}_{AB}^{(-)},$ where $\sigma_{AB}=\frac{1}{2}(U_{A}\rho_{AB}U_{A}^\dagger + V_{B}\rho_{AB}V_{B}^\dagger),$ and $\tilde{\sigma}_{AB}^{(-)}=\frac{1}{4}(U_{A}-V_{B})\rho_{AB}(U_{A}-V_{B})^\dagger.$ Since $T(\tilde{\sigma}_{AB},\rho_{AB}) = \frac{1}{2}||\tilde{\sigma}_{AB}-\rho_{AB}||_{1},$ we can apply the triangle inequality $||A+B||_1\leq||A||_1+||B||_1,$ to get $T(\tilde{\sigma}_{AB},\rho_{AB}) \leq T(\sigma_{AB},\rho_{AB})+\frac{1}{2}\text{Tr}(\tilde{\sigma}_{AB}^{(-)}).$ Using this with Eq.~(\ref{Eq:MaxWinProbAltForm}), and noting that $\text{Tr}(\tilde{\sigma}_{AB}^{(-)}) + \text{Tr}(\tilde{\sigma}_{AB})=1,$ we arrive at $p_{\text{pnp}}(\rho_{AB})\leq \frac{1}{2}+\frac{1}{2}T(\sigma_{AB},\rho_{AB}).$

Before progressing further, we can understand this more intuitively by considering a variation on the game, where C no longer performs a measurement afterwards to see if they have decohered the state. This means we are just focusing on A and B's ability to record the presence of a particle. The win probability for this game is derived similarly to before, and we find
\begin{equation*}
    \tilde{p}_{\text{pnp}}(\rho_{AB})  = \frac{1}{2} + \frac{1}{2}T\big(\rho_{AB},\frac{U_A\rho_{AB}U_{A}^\dagger + V_B\rho_{AB}V_{B}^\dagger}{2}\big),
\end{equation*}
where we have already performed the maximization over the choice of POVMs.

This is an easier game by construction, so the maximum win probability for it must upper bound the maximum win probability of the original game. This therefore gives the same bound that we arrived at via the triangle inequality.

In order to continue we make use of the following inequalities
\begin{equation}\label{Eq:QTraceIneqs}
\begin{split}
    \max_{U_A,V_B}T\big(\rho_{AB},\frac{U_A\rho_{AB}U_{A}^\dagger + V_B\rho_{AB}V_{B}^\dagger}{2}\big)
    & \leq \frac{1}{2}\max_{U_A,V_B}(T(\rho_{AB},U_A\rho_{AB}U_{A}^\dagger) + T(\rho_{AB},V_B\rho_{AB}V_{B}^\dagger)), \\
    & \leq \max_{U_{AB}}T(\rho_{AB},U_{AB}\rho_{AB}U_{AB}^\dagger).
\end{split}
\end{equation}
Here we have used the convexity of the trace distance \cite{nielsen2002quantum} and the fact that a maximization over all possible global unitaries will always give a value greater than or equal to that obtained by maximization over locally restricted unitaries.

We now make use of the following Lemma.
\begin{lemma}
For two density matrices $\rho$ and $\sigma$ defined on the same Hilbert space $\mathcal{H}$ of dimension $n$. It holds that
\begin{equation*}
    T(\rho,\sigma) \leq T_{c}(\lambda^{\uparrow},\mu^{\downarrow}),
\end{equation*}
where $T$ is the quantum trace distance, $T_{c}$ the Kolmogorov (classical trace) distance, $\lambda^{\uparrow}$ is the vector of $n$ eigenvalues of $\rho$ arranged in ascending order and $\mu^{\downarrow}$ is the vector of $n$ eigenvalues of $\sigma$ arranged in descending order, where these vectors of eigenvalues include any zero values.
\end{lemma}

To prove this we start by writing $\rho=\sum_i \lambda_i |\psi_i\rangle\langle\psi_i|,$ $\sigma=\sum_i \mu_i |\phi_i\rangle\langle\phi_i|,$ and $\rho-\sigma = \sum_i a_i |\alpha_i\rangle\langle\alpha_i|,$ with all the eigenvalues arranged in ascending order such that $\lambda_1\leq\lambda_2\leq...\leq\lambda_n$ and similarly for the others. The trace distance $T(\rho,\sigma)=\frac{1}{2}\text{Tr}|\rho-\sigma|,$ is given by the sum of the positive eigenvalues of $\rho - \sigma,$ i.e. $T(\rho,\sigma)=\sum_{i,a_i\geq0}a_i.$

We now shall make use of the Min-max theorem which we state as follows. Consider a Hermitian operator $H,$ defined on a Hilbert space $\mathcal{H}$ of dimension $n.$ We denote $k$ dimensional subspaces as $\mathcal{H}_{k},$ i.e. this denotes any Hilbert space that satisfies  $\mathcal{H}_{k}\subseteq\mathcal{H},$ and $\text{dim}(\mathcal{H}_{k})=k.$ Now working with normalised vectors $|\langle\chi|\chi\rangle|=1,$ the Min-max theorem states that the eigenvalues of $H,$ arranged such that $h_1\leq h_2\leq...\leq h_n,$ satisfy 
\begin{equation*}
    h_k = \min_{\mathcal{H}_{k}} \max_{|\chi\rangle\in\mathcal{H}_{k}} \langle\chi|H|\chi\rangle.
\end{equation*}
A simple corollary of this is that 
\begin{equation*}
    h_k \leq h_{k+l} \leq \max_{|\chi\rangle\in\mathcal{H}_{k+l}} \langle\chi|H|\chi\rangle.
\end{equation*}
We apply this Corollary to $\rho-\sigma$ to get
\begin{equation}\label{Eq:akBound}
    a_k \leq \max_{|\chi\rangle\in\mathcal{H}_{k+l}} \sum_{i=1}^{n} (\lambda_i |\langle\chi|\psi_i\rangle|^2 - \mu_i |\langle\chi|\phi_i\rangle|^2).
\end{equation}
This is true for any choice of $\mathcal{H}_{k+l}.$ We shall proceed by defining a particular case. To this end we define the linear operator
\begin{equation*}
    L = \sum_{i=1}^{n-k} |e_i\rangle (\sqrt{\lambda_{n-i+1}} \langle\psi_{n-i+1}| + \sqrt{\mu_{i}} \langle\phi_i|),
\end{equation*}
where the $|e_i\rangle$ form some set of orthonormal vectors.

We now consider the kernel of $L.$ We know that the kernel of a linear operator is a vector space, and $L|\chi\rangle=0$ implies the following $n-k$ constraints
\begin{equation}\label{Eq:Constraints}
     \sqrt{\lambda_{n-i+1}} \langle\psi_{n-i+1}|\chi\rangle + \sqrt{\mu_{i}} \langle\phi_i|\chi\rangle = 0.
\end{equation}
This specifies a $k+l$ dimensional subspace with $l=0$ if the constraints are independent and $l\neq0$ otherwise. From this we see that we can choose to define $\mathcal{H}_{k+l}$ as the kernel of $L.$

We now rewrite Eq.~(\ref{Eq:akBound}) as
\begin{equation*}
    a_k \leq \max_{|\chi\rangle\in\mathcal{H}_{k+l}} \bigg( \sum_{i=1}^{k} \lambda_i |\langle\chi|\psi_i\rangle|^2 - \sum_{i=n-k+1}^{n}\mu_i |\langle\chi|\phi_i\rangle|^2 
    + \sum_{i=1}^{n-k} (\lambda_{n-i+1} |\langle\psi_{n-i+1}|\chi\rangle|^2 - \mu_i |\langle\phi_i|\chi\rangle|^2 \bigg).
\end{equation*}
Substituting in for the constraints of Eq.~(\ref{Eq:Constraints}) we have
\begin{equation*}
    a_k \leq \max_{|\chi\rangle} \big( \sum_{i=1}^{k} \lambda_i |\langle\chi|\psi_i\rangle|^2 - \sum_{i=n-k+1}^{n}\mu_i |\langle\chi|\phi_i\rangle|^2 \big).
\end{equation*}
From this we find
\begin{equation*}
    a_k \leq  \lambda_k - \mu_{n-k+1}.
\end{equation*}
Now we use this to write
\begin{equation*}
\begin{split}
    T(\rho,\sigma) & = \sum_{i,a_i\geq0}a_i, \\
    & \leq \sum_{i,a_i\geq0}(\lambda_i - \mu_{n-i+1}), \\
    & \leq \sum_{i,(\lambda_i - \mu_{n-i+1})\geq0}(\lambda_i - \mu_{n-i+1}), \\
    & = \sum_{j,(\lambda_{n-j} - \mu_{j+1})\geq0}(\lambda_{n-j} - \mu_{j+1}).
\end{split}
\end{equation*}
By writing the last term in this way it is now clear that this is equal to the Kolmogorov (classical trace) distance $T_{c}(\lambda^{\uparrow},\mu^{\downarrow}),$ therefore we have the stated result.

Now since $\rho_{AB}$ and $U_{AB}\rho_{AB}U^\dagger_{AB}$ have the same eigenvalues we have that $T(\rho_{AB},U_{AB}\rho_{AB}U_{AB}^\dagger)\leq T_{c}(\lambda^{\uparrow},\lambda^{\downarrow}),$ and using this with Eq.~(\ref{Eq:QTraceIneqs}) we arrive at the final form for the record quality bound
\begin{equation*}
    p_{\text{pnp}}^{\text{m}}(\rho_{AB}) \leq \frac{1}{2}+\frac{1}{2}T_{c}(\lambda^{\uparrow},\lambda^{\downarrow}).
\end{equation*}

\begin{figure}
\includegraphics[width=0.7\linewidth]{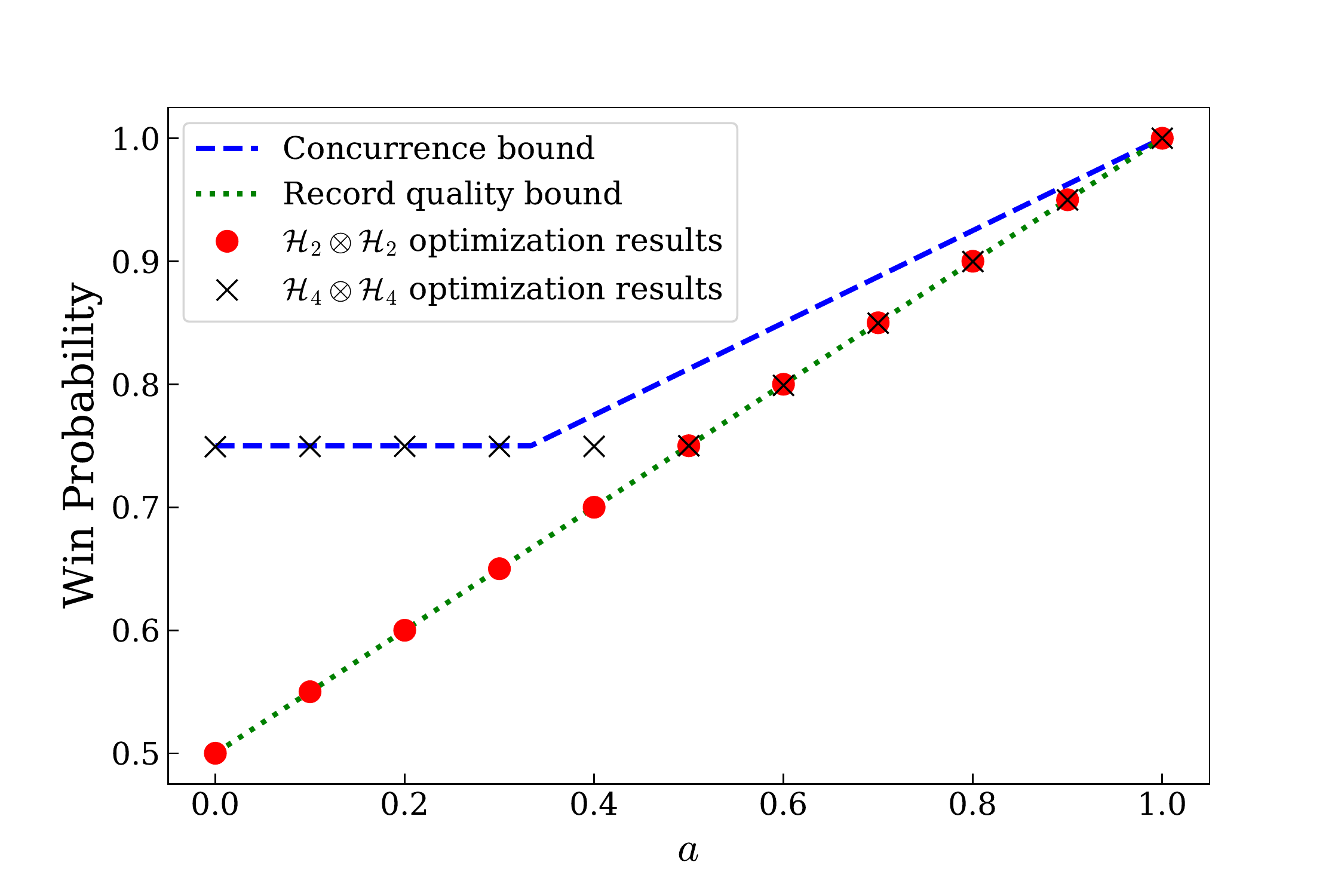}
\caption{Plot of numerically optimized PNP win probabilities and bounds for 2-qubit states of the form $a|\psi^{k}\rangle\langle\psi^{k}|+\frac{1-a}{4}\mathds{1},$ for different values of $a.$ This is as presented in the main text, only here we also have data points for the case where A and B are granted access to additional pure classical resources, denoted by the crosses.}
\label{fig:WernerState}
\end{figure}

As stated in the main text, this bound can be saturated for certain mixed states such as the Werner-like states. However, it is worth noting that with a modification to the rules of the game, this could in principle be circumvented. If we allow A and B unlimited access to additional pure separable resources then the bound should no longer apply. This introduction of additional classical resources is seen to be equivalent to embedding in a higher dimensional Hilbert space, which then zero-pads the vector of eigenvalues such that $T_{c}(\lambda^{\uparrow},\lambda^{\downarrow})=0.$ If A and B share a Werner state and are given access to the additional pure qubit state $|00\rangle_{A'B'}\langle00|,$ this is equivalent to allowing them to optimize their unitaries $U_{A},V_{B}$ over the group $U(4)$ as opposed to the previous case where we used $U(2).$ Figure~\ref{fig:WernerState} also illustrates the results for this. As expected the win probability is never significantly below the $0.75$ classical limit. However we note the striking feature that the win probability still does not get above the classical limit until $a>\frac{1}{2},$ and that when it does it follows the $\mathcal{H}_2\otimes\mathcal{H}_2$ record quality bound.

\subsection{PNP mixtures of two Bell states}\label{appen:MixBellStates}
 Here we show that mixtures of two Bell states can saturate the concurrence bound. Taking the specific example $\rho=a|\psi^{+}\rangle\langle\psi^{+}|+(1-a)|\psi^{-}\rangle\langle\psi^{-}|,$ we find $\tilde{\rho}=(Y\otimes Y)\rho^{*}(Y\otimes Y)=\rho,$ so $\rho\tilde{\rho}=\rho^2,$ and from this we find $C(\rho)=1-2a$ for $a\leq\frac{1}{2},$ and $C(\rho)=2a-1$ for $a\geq\frac{1}{2}.$ Now for $a\leq\frac{1}{2},$ we calculate the win probability with the choices $U_A=X_A,$ and $V_B=-X_B,$ noting that this choice maps $\rho$ to an orthogonal state. For this choice of unitaries we find $p_\text{w}(\rho) = 1-\frac{a}{2}=\frac{3}{4}+\frac{1-2a}{4},$ and for $a\geq\frac{1}{2},$ we take $U_A=X_A,$ and $V_B=X_B,$ for which we get $p_\text{w}(\rho) = \frac{1}{2}(1+a)=\frac{3}{4}+\frac{2a-1}{4}.$ We see that in both instances we are exactly saturating the concurrence bound.

\subsection{BD game concurrence bound}\label{appen:ConBoundBDG}

The win probability for a pure state $|\psi\rangle,$ can be written as
\begin{equation*}
    p_{\text{bd}} = \frac{k_{11}+k_{22}+\sqrt{(k_{11}+k_{22})^2 - 4 k_{12}k_{21}}}{4}.
\end{equation*}
where $k_{ij}=\langle\psi|K_{i}^\dagger K_{j}|\psi\rangle,$ with $K_1 = (U_A+V_B)/2$ and $K_1 = (U_A V_B+\mathds{1})/2$

One can rewrite the terms as
\begin{equation*}
    k_{11}+k_{22} = 1 + \langle\psi|AB|\psi\rangle,
\end{equation*}
\begin{equation*}
    k_{12} = \frac{1}{2}\langle\psi|(A+B)|\psi\rangle,
\end{equation*}
where we have defined the local operators $A=(U_{A}+U_{A}^\dagger)/2,$ and $B=(V_{B}+V_{B}^\dagger)/2.$

We can then re-express the term under the square root as
\begin{equation*}
\begin{split}
    (1+\langle\psi|AB|\psi\rangle)^2-\langle\psi|(A+B)|\psi\rangle^2 &= \langle\psi|(\mathds{1}+AB)|\psi\rangle^2-\langle\psi|(A+B)|\psi\rangle^2, \\
    &= \langle\psi|(\mathds{1}+AB+A+B)|\psi\rangle\langle\psi|(\mathds{1}+AB-A-B)|\psi\rangle, \\
    &= 2^{4}\langle\psi|\Pi_{A}^{+}\Pi_{B}^{+}[|\psi\rangle\langle\psi|\Pi_{A}^{-}\Pi_{B}^{-}|\psi\rangle.
\end{split}
\end{equation*}
where we define the positive semi-definite operators $\Pi_{A}^{\pm} = (\mathds{1}\pm A)/2,$ and similarly for $B.$

We also write
\begin{equation*}
\begin{split}
    k_{11}+k_{22} &= \langle\psi|(\mathds{1}+AB)|\psi\rangle, \\
    &= \frac{1}{2}\langle\psi|(\mathds{1}+A)(\mathds{1}+B)-(\mathds{1}-A)(\mathds{1}-B)|\psi\rangle, \\
    &= 2(\langle\psi|\Pi_{A}^{+}\Pi_{B}^{+}|\psi\rangle + \langle\psi|\Pi_{A}^{-}\Pi_{B}^{-}|\psi\rangle).
\end{split}
\end{equation*}

Putting these together we have
\begin{equation*}
\begin{split}
    p_{\text{bd}} &= \frac{\langle\psi|\Pi_{A}^{+}\Pi_{B}^{+}|\psi\rangle + \langle\psi|\Pi_{A}^{-}\Pi_{B}^{-}|\psi\rangle+2\sqrt{\langle\psi|\Pi_{A}^{+}\Pi_{B}^{+}|\psi\rangle\langle\psi|\Pi_{A}^{-}\Pi_{B}^{-}|\psi\rangle}}{2}, \\
    &= \frac{1}{2}\big(\sqrt{\langle\psi|\Pi_{A}^{+}\Pi_{B}^{+}|\psi\rangle} + \sqrt{\langle\psi|\Pi_{A}^{-}\Pi_{B}^{-}|\psi\rangle}\big)^2.
\end{split}
\end{equation*}

Now considering qubit states we write $|\psi\rangle = \sqrt{r} |00\rangle + \sqrt{1-r}|11\rangle,$ and we denote $a_{i,j}=\langle i|\Pi_{A}^{+}|j\rangle,$ and $b_{i,j}=\langle i|\Pi_{B}^{+}|j\rangle.$ Since $\Pi_{A}^{+}\geq 0,$ we have $a_{00}a_{11}\geq|a_{01}|^2,$ and since $\mathds{1}-\Pi_{A}^{+}\geq 0,$ we have $(1-a_{00})(1-a_{11})\geq|a_{01}|^2.$ Using these we write
\begin{equation*}
\begin{split}
    \langle\psi|\Pi_{A}^{+}\Pi_{B}^{+}|\psi\rangle & \leq r a_{00} b_{00} + (1-r) a_{11} b_{11} + 2\sqrt{r(1-r)}\sqrt{a_{00}a_{11}b_{00}b_{11}}, \\
    &= (\sqrt{r a_{00}b_{00}} + \sqrt{(1-r) a_{11}b_{11}})^2,
\end{split}
\end{equation*}
and similarly
\begin{equation*}
    \langle\psi|\Pi_{A}^{-}\Pi_{B}^{-}|\psi\rangle \leq 
    = (\sqrt{r (1-a_{00})(1-b_{00})} + \sqrt{(1-r) (1-a_{11})(1-b_{11})})^2.
\end{equation*}
This now gives us that
\begin{equation*}
\begin{split}
    p_{\text{bd}}^{\text{m}} &\leq \frac{1}{2}[\sqrt{r}(\sqrt{a_{00}b_{00}} + \sqrt{(1-a_{00})(1-b_{00})} + \sqrt{1-r}(\sqrt{a_{11}b_{11}} + \sqrt{(1-a_{11})(1-b_{11})})]^2,\\
    &\leq \frac{1}{2}[\sqrt{r} + \sqrt{1-r}]^2, \\
    &= \frac{1}{2}[1 + C(|\psi\rangle)].
\end{split}
\end{equation*}

The bit flip strategy saturates this bound thus proving equality. We can then extend to mixed states in the usual manner to obtain
\begin{equation*}
    p_{\text{bd}}^{\text{m}}(\rho_{AB})\leq \frac{1}{2} + \frac{1}{2}C(\rho_{AB}),
\end{equation*}

\subsection{Demonstration on IBM quantum processor}\label{appen:Demo}

We implemented demonstrations for both the PNP and BD game, using different circuits to represent the various cases of different states prepared by Charlie. For both games we ran circuits for Alice and Bob having an entangled resource state $|\psi^{+}\rangle$ and only having a separable resource state $|00\rangle.$ We designed the demonstrations to be implemented on a sequence of four qubits with linear connectivity. This is because it is a simple approach that requires only low depth circuits and is consistent with the qubit connectivity of the IBM device we selected, but naturally one could create more complex circuits using other architectures. We used \emph{Paris} as it was the most recent device provided by IBM at the time of running.

\begin{figure}[h]
\begin{center}
\includegraphics[scale=0.55]{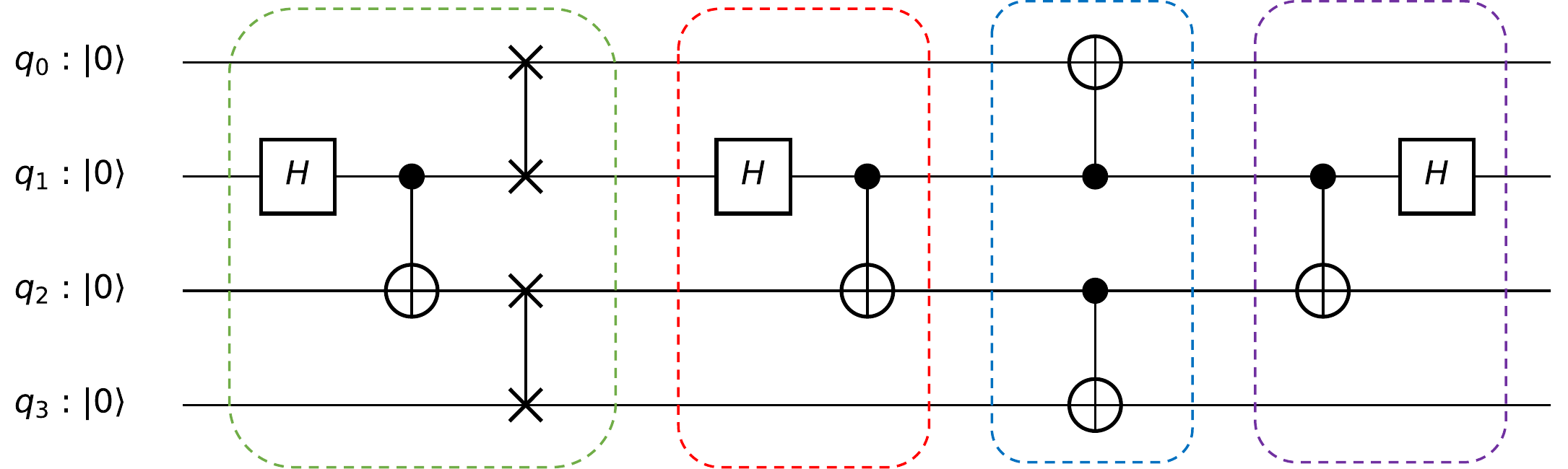}

\vspace{0.8cm}

\includegraphics[scale=0.55]{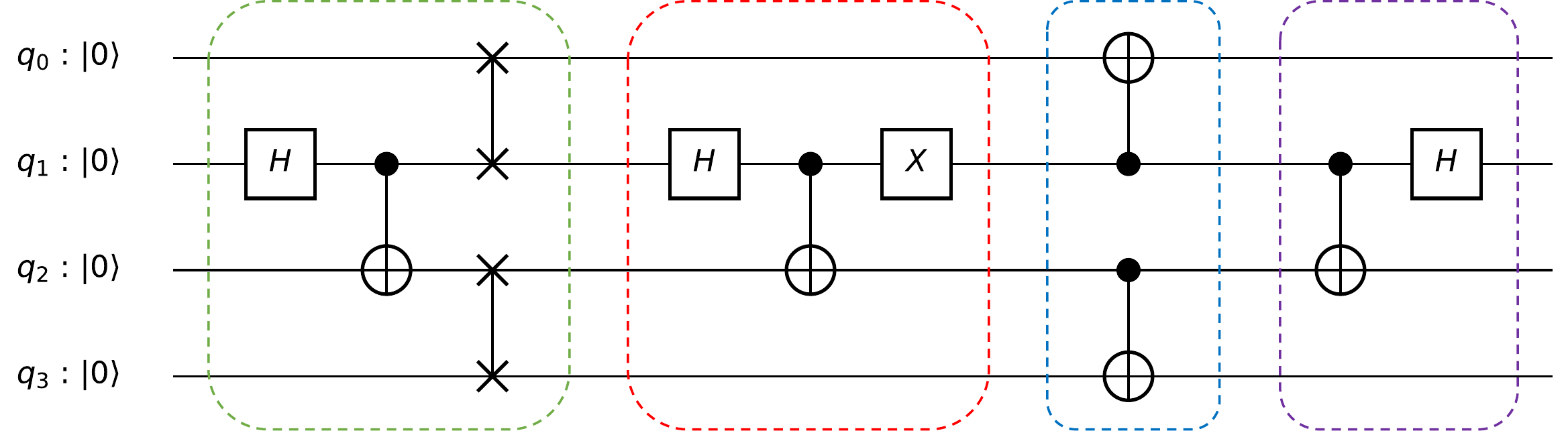}

\vspace{0.8cm}

\includegraphics[scale=0.55]{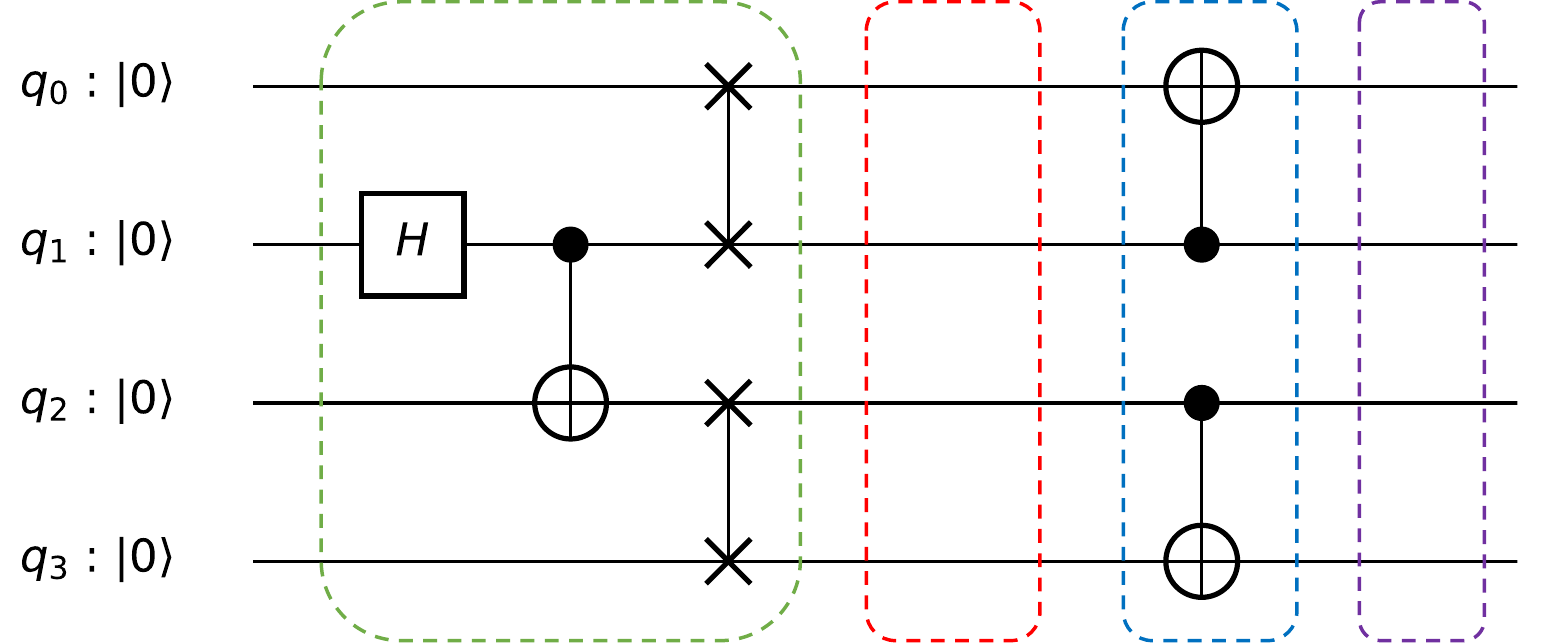}

\vspace{0.1cm}

\caption{Circuit diagrams for A and B using the entangled resource state $|\phi^{+}\rangle$ and Charlie sending $|\phi^{+}\rangle,$ $|\psi^{+}\rangle,$ and $|00\rangle,$ respectively. The circuits are partitioned into sections: AB State Preparation, C State Preparation, Interaction, and Measurement.}\label{Fig:CircuitDiagram}
\end{center}
\end{figure}

The circuits for an initial entangled resource state are illustrated in Fig.~\ref{Fig:CircuitDiagram}. The circuits represent the cases where Charlie prepares $|\phi^{+}\rangle,|\psi^{+}\rangle,$ and $|00\rangle$ respectively. The qubits $q_0,q_3$ are used to represent the resource state of Alice and Bob so correspond to $A,B$ respectively. The qubits $q_1,q_2$ will represent the state Charlie sends and correspond to $A_p,B_p,$ where this labelling ensures that $A$ interacts with $A_p$ and $B$ with $B_p.$ However, qubits $q_1,q_2$ have an extra use as before representing $A_p,B_p,$ they will be used to distribute the initial entanglement between $q_0$ and $q_3.$ We shall clarify the action of the circuits by explicitly describing their four stages.

The first stage is to prepare $q_0,q_3$ in the $|\psi^{+}\rangle_{q_0,q_3}$ Bell state, as this will be the entangled resource state. This is achieved by first preparing $q_1,q_2$ in the state $|\psi^{+}\rangle{q_1,q_2}$ by applying $\text{H}_{q_{1}}$ followed by $\text{CX}_{q_{1}q_{2}},$ and then moving out this state by applying the swaps $\text{SWAP}_{q_{1}q_{0}}$ and $\text{SWAP}_{q_{2}q_{3}}.$ This stage is the same for all three of the entangled cases, and when we do the separable cases we simply omit it and thereby Alice and Bob start with the unentangled $|00\rangle_{q_0,q_3}$

The second stage is to prepare Charlie's question state. This can be one of the three states $|\phi^{+}\rangle_{q_1,q_2},|\psi^{+}\rangle_{q_1,q_2},|00\rangle_{q_1,q_2},$ where for the BD game we use the first two and for the PNP game we use the latter two. The $|\phi^{+}\rangle$ state is prepared by applying $\text{CX}_{q_{1}q_{2}}\text{H}_{q_{1}},$ the $|\psi^{+}\rangle$ state by applying $\text{X}_{q_1}\text{CX}_{q_{1}q_{2}}\text{H}_{q_{1}},$ and the $|00\rangle$ state requires no operations. This is the same for the separable cases because Alice and Bob's tactics have no bearing on the game as administered by Charlie.

The third stage is the interaction stage and is the same in all cases. It is made up of the controlled unitaries $\text{CX}_{q_{1}q_{0}}$ and $\text{CX}_{q_{2}q_{3}}.$ This is followed by the fourth and final stage which is the measurement stage. If Charlie has prepared an entangled state then he applies $\text{CX}_{q_{1}q_{2}}$ followed by $\text{H}_{q_{1}},$ and then all qubits are measured in the computational basis.

\begin{figure}
\begin{center}
\includegraphics[width=.85\linewidth]{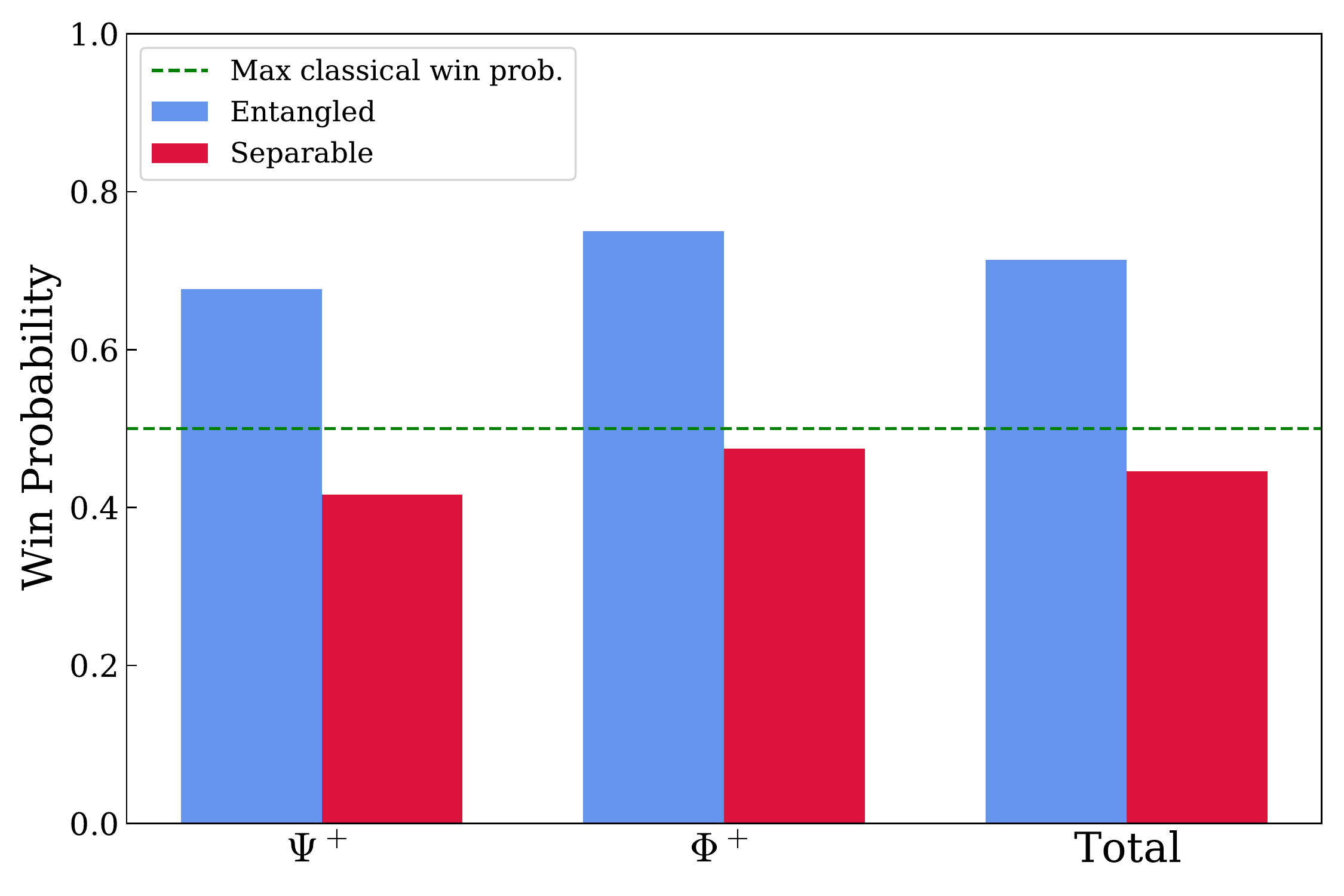}

\vspace{0.8cm}

\includegraphics[width=.85\linewidth]{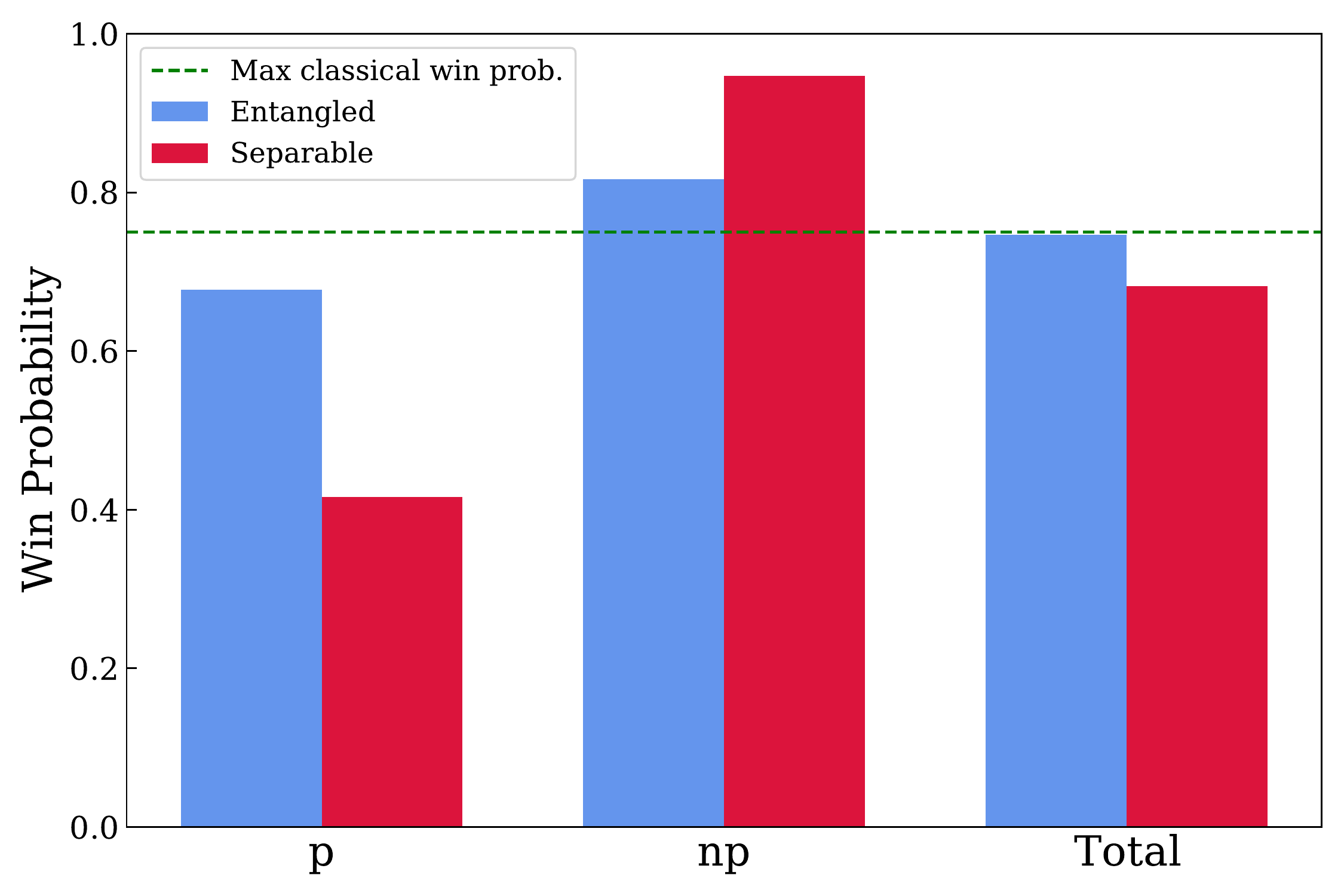}

\vspace{0.1cm}

\caption{Plot of the results calculated from the Paris device measurements. The top plot is for the BD game and the bottom plot is for the PNP game. The total win probability is calculated for equal probability of sending either state. The blue bars are for the entangled initial resource state, the red for the separable, and the green line is to show the maximum classical win probabilities of $\frac{1}{2},$ and $\frac{3}{4},$ respectively. }\label{Fig:ParisResultsPlots}

\end{center}
\end{figure}

Each circuit was run with 8192 shots, with the win probabilities calculated from the measurement results, and these are presented in Fig.~\ref{Fig:ParisResultsPlots}. Focusing first on the BD results, we see that the entangled state total win probability is higher than the classical limit of $0.5,$ from which we calculate that it demonstrates a usable concurrence of $0.42,$ and thus a convincing delocalised interaction. However, the realised win probability is far below the ideal case of $1.$ Additionally, we note the separable case is below the classical limit, demonstrating the problem with noise even for the classical strategy. For the standard PNP game we note that we were unable to convincingly demonstrate non-classical performance, as the entangled state total win probability approximately matches the classical limit of $0.75,$ it does not exceed it. Despite this, it should be noted that the entangled strategy still out performs the separable strategy as run on the device, which indicates the entanglement is still acting in beneficially. Furthermore, if we use the different sending probabilities $P_{\text{p}}=\frac{2}{3},$ $P_{\text{np}}=\frac{1}{3}$ (see next section) then we get total win probability $0.72$ which is higher than this game's classical limit of $\frac{2}{3}.$ Note however, that we currently do not have a way to directly relate this violation to an entanglement measure. Clearly the noise in the device is significant, but we still see non-classical behaviour and with the impressive rate of improvement in this field~\cite{kjaergaard2019superconducting} one can envisage even better results in the near future. 

Though the BD case apparently clearly demonstrates delocalised-interactions, as in Bell tests, there are a number of loopholes~\cite{larsson2014loopholes} one could consider to avoid the conclusion that a delocalised interaction took place. Chief amongst them is that A and B actually interacted during the game, which could be solved by keeping them space-like separated for the duration. The difficulty with this comes from having to reliably and quickly send quantum information, which for superconducting qubits is not currently feasible, therefore photonic qubits~\cite{kok2007linear} could prove to be more appropriate.

\subsection{PNP with different sending probabilities}\label{appen:G3DifProbs}
Accounting for different probabilities for C sending a particle $P_{\text{p}}$ and no particle $P_{\text{np}},$ the win probability can be written as
\begin{equation*}
    p_{\text{pnp}}(\rho_{AB}) = P_{\text{np}} + \text{eigs}_{+}(P_{\text{p}}\tilde{\sigma}_{AB}-P_{\text{np}}\rho_{AB}),
\end{equation*}
For pure states, using the same logic and notation as before via Eq.~(\ref{Eq:generalEigenVal}) we arrive at needing to maximise
\begin{equation*}
    n = P_{\text{p}}k - P_{\text{np}} + \sqrt{(P_{\text{p}}k+P_{\text{np}})^2-4P_{\text{p}}P_{\text{np}}\kappa}
\end{equation*}
As before we instead maximise an upper bound on this given by
\begin{equation*}
    \tilde{n} = P_{\text{p}}k_{b} - P_{\text{np}} + \sqrt{(P_{\text{p}}k_{b}+P_{\text{np}})^2-4P_{\text{p}}P_{\text{np}}\kappa},
\end{equation*}
where we have $k_{b} = \frac{1}{2}[1+ uv\cos\Delta\phi + G(|\psi\rangle)],$ and $\kappa=\frac{1}{4}(u^2+v^2+2uv\cos\Delta\phi).$
Calculating the partial derivatives we find
\begin{equation*}
    \frac{\partial \tilde{n}}{\partial\Delta\phi} =-\frac{1}{2}uv\sin\Delta\phi
    (P_{\text{p}} +\frac{P_{\text{p}}^{2}k_{b}- P_{\text{p}}P_{\text{np}}}{\sqrt{(P_{\text{p}}k_{b}+P_{\text{np}})^2-4P_{\text{p}}P_{\text{np}}\kappa}}),
\end{equation*}
\begin{equation*}
    \frac{\partial \tilde{n}}{\partial u} = \frac{1}{2}(P_{\text{p}} +\frac{P_{\text{p}}^{2}k_{b}- P_{\text{p}}P_{\text{np}}}{\sqrt{(P_{\text{p}}k_{b}+P_{\text{np}})^2-4P_{\text{p}}P_{\text{np}}\kappa}})(v\cos\Delta\phi ) - \frac{P_{\text{p}}P_{\text{np}}}{\sqrt{(P_{\text{p}}k_{b}+P_{\text{np}})^2-4P_{\text{p}}P_{\text{np}}\kappa}}u,
\end{equation*}
\begin{equation*}
    \frac{\partial \tilde{n}}{\partial v} = \frac{1}{2}(P_{\text{p}} +\frac{P_{\text{p}}^{2}k_{b}- P_{\text{p}}P_{\text{np}}}{\sqrt{(P_{\text{p}}k_{b}+P_{\text{np}})^2-4P_{\text{p}}P_{\text{np}}\kappa}})(v\cos\Delta\phi ) - \frac{P_{\text{p}}P_{\text{np}}}{\sqrt{(P_{\text{p}}k_{b}+P_{\text{np}})^2-4P_{\text{p}}P_{\text{np}}\kappa}}v,
\end{equation*}
Setting these expressions equal to zero we obtain the three equations
\begin{equation}\label{Eq:newpd1}
    0 = \tilde{X}uv\sin\Delta\phi,
\end{equation}
\begin{equation}\label{Eq:newpd2}
    0 = \tilde{X}v\cos\Delta\phi - \tilde{Y}u,
\end{equation}
\begin{equation}\label{Eq:newpd3}
    0 = \tilde{X}u\cos\Delta\phi - \tilde{Y}v,
\end{equation}
where $\tilde{X}\equiv(P_{\text{p}} +\frac{P_{\text{p}}^{2}k_{b}- P_{\text{p}}P_{\text{np}}}{\sqrt{(P_{\text{p}}k_{b}+P_{\text{np}})^2-4P_{\text{p}}P_{\text{np}}\kappa}})$ and $\tilde{Y}=\frac{2P_{\text{p}}P_{\text{np}}}{\sqrt{(P_{\text{p}}k_{b}+P_{\text{np}})^2-4P_{\text{p}}P_{\text{np}}\kappa}}.$ Note that setting $P_{\text{p}}=P_{\text{np}}=1/2,$ recovers the equations we had before.

For $P_{\text{p}}\neq0$ and $P_{\text{np}}\neq0,$ we as before find that neither $\tilde{X}$ nor $\tilde{Y}$ can equal zero. It follows that to satisfy Eq.~(\ref{Eq:newpd1}) we must take $u=0,$ $v=0,$ or $\sin\Delta\phi=0.$ If we consider $u=0$ then Eq.~(\ref{Eq:newpd3}) implies $v=0.$ Similarly if we consider $v=0,$ then Eq.~(\ref{Eq:newpd2}) implies $u=0.$ So in both cases we have $u=v=0,$ and this naturally makes the choice of $\Delta\phi$ irrelevant. Therefore the only choice we now have to consider is $\sin\Delta\phi=0.$

Taking this case we write Eq.~(\ref{Eq:newpd2}) and Eq.~(\ref{Eq:newpd3}) as $\pm \tilde{X}v=\tilde{Y}u$ and $\pm \tilde{X}u=\tilde{Y}v$ respectively. Now if either $\pm \tilde{X}v=\tilde{Y}u=0$ or $\pm \tilde{X}u=\tilde{Y}v=0,$ then this implies $u=v=0,$ so we just need to consider the option of $\pm \tilde{X}v=\tilde{Y}u\neq0$ and $\pm \tilde{X}u=\tilde{Y}v\neq0.$ In this case we can divide through and get $u^2=v^2$ so $u=\pm v.$ Substituting back we have $(\tilde{X}\pm \tilde{Y})u=0.$ We have
\begin{equation*}
    \tilde{X}\pm \tilde{Y} = P_{\text{p}} +\frac{P_{\text{p}}^{2}k_{b}- P_{\text{p}}P_{\text{np}} \pm 2P_{\text{p}}P_{\text{np}} }{\sqrt{(P_{\text{p}}k_{b}+P_{\text{np}})^2-4P_{\text{p}}P_{\text{np}}\kappa}},
\end{equation*}
and see that for the relevant ranges only $\tilde{X}-\tilde{Y}$ could be equal to zero so we write
\begin{equation*}
    P_{\text{p}} +\frac{P_{\text{p}}^{2}k_{b}- 3P_{\text{p}}P_{\text{np}} }{\sqrt{(P_{\text{p}}k_{b}+P_{\text{np}})^2-4P_{\text{p}}P_{\text{np}}\kappa}}=0.
\end{equation*}
The solution to this equation is $k_{b}=\frac{P_{\text{np}}}{P_{\text{p}}} + \frac{\kappa}{2}.$ Plugging in the definitions of $k_{b}$ and $\kappa$ and using that $\sin\Delta\phi=0,$ and $u^2=v^2,$ we have
\begin{equation*}
    u^2\mp u^2 = 2(1+G(|\psi\rangle)) - \frac{4P_{\text{np}}}{P_{\text{p}}}.
\end{equation*}
We can check for our previous result by setting $P_{\text{np}}=P_{\text{p}}=\frac{1}{2},$ we then have that the right hand side is always negative for $G(|\psi\rangle)<1,$ therefore there is no real solution. For $P_{\text{np}}\geq P_{\text{p}}$ this argument holds, but for $P_{\text{np}}< P_{\text{p}}$ things are more complicated. However, we can extract a useful extra result for the case of separable states $G(|\psi\rangle)=0.$ For these we find that the right hand side is negative for $P_{\text{p}}<\frac{2}{3},$ and for this by similar arguments to before we have that the turning point is given by $u=v=0.$ This gives the maximum separable win probability in this range as $P_{\text{np}}+\frac{1}{2}P_{\text{p}}.$ It makes sense that this regime breaks down at $P_{\text{p}}=\frac{2}{3}$ because this is when doing nothing and simply always guessing that there is a particle, will give equal win probability to measuring all the time and losing half the time when there is a particle.

\subsection{Trace distance inequality}\label{appen:TDProof}
First we prove that for the special case of pure separable states (product states) we have
\begin{equation}\label{Eq:ProductStateTraceDist}
    T(U_{A}|\psi\rangle_{AB},|\psi\rangle_{AB})\leq T(U_{A}|\psi\rangle_{AB},V_{B}|\psi\rangle_{AB}).
\end{equation}
To prove this we first note that for product states $|\langle \psi|V_{B}^\dagger U_{A}|\psi\rangle| = |\langle \psi|V_{B}^\dagger|\psi\rangle\langle \psi| U_{A}|\psi\rangle|.$ It then follows that
\begin{equation*}
\begin{split}
    T(U_{A}|\psi\rangle_{AB},V_{B}|\psi\rangle_{AB}) & =
    \sqrt{1-|\langle \psi|V_{B}^\dagger U_{A}|\psi\rangle|^2} \\
    & =\sqrt{1-|\langle \psi|V_{B}^\dagger|\psi\rangle\langle \psi| U_{A}|\psi\rangle|^2} \\
    & \geq \sqrt{1-|\langle\psi| U_{A}|\psi\rangle|^2} \\
    & = T(U_{A}|\psi\rangle_{AB},|\psi\rangle_{AB})
\end{split}
\end{equation*}
We use this result to prove the general statement as follows. Using the fact that a separable state can be decomposed into a convex mixture of product states $\rho_{AB}=\sum_{i}q_{i}|\psi_{i}\rangle_{AB}\langle\psi_{i}|,$ we write
\begin{equation*}
\begin{split}
    T(U_{A}\rho_{AB}U_{A}^\dagger,\rho_{AB}) & =
    T(\sum_{i}q_{i}U_{A}|\psi_{i}\rangle_{AB}\langle\psi_{i}|U_{A}^\dagger , \sum_{i}q_{i}|\psi_{i}\rangle_{AB}\langle\psi_{i}|) \\
    & \leq \sum_{i}q_{i}T(U_{A}|\psi_{i}\rangle_{AB}, |\psi_{i}\rangle_{AB})  \\
    & \leq \sum_{i}q_{i}T(U_{A}|\psi_{i}\rangle_{AB} , V_{B}|\psi_{i}\rangle_{AB}) \\
    & \leq T(U_{A}|\Psi\rangle_{ABC},V_{B}|\Psi\rangle_{ABC}).
\end{split}
\end{equation*}
To get to the second line we use convexity of the trace distance and to get to the third we use Eq.~(\ref{Eq:ProductStateTraceDist}). To arrive at the last line we consider doing state discrimination between the two states $U_{A}|\Psi\rangle_{ABC},V_{B}|\Psi\rangle_{ABC}.$ All purifications of $\rho_{AB}=\sum_{i}q_{i}|\psi_{i}\rangle\langle\psi_{i}|$ can be written as $|\Psi\rangle_{ABC} = \mathcal{U}_{C'\rightarrow C}\sum_{i}|\psi_{i}\rangle_{AB}|i\rangle_{C'},$ where $\mathcal{U}_{C'\rightarrow C}$ is an isometry.  A strategy to discriminate between $U_{A}|\Psi\rangle_{ABC},$ and $V_{B}|\Psi\rangle_{ABC},$ is given by first undoing any isometry, then measuring in the $|i\rangle_{C}$ basis, and finally performing the optimally discriminating measurement between $U_{A}|\psi_{i}\rangle_{AB}$ and $V_{B}|\psi_{i}\rangle_{AB}.$ This will correctly discriminate between the two states with probability $\sum_i \frac{q_{i}}{2}(1+T(U_{A}|\psi_{i}\rangle_{AB},V_{B}|\psi_{i}\rangle_{AB}),$ however the maximal discrimination probability is given by $\frac{1}{2}(1+T(U_{A}|\Psi\rangle_{ABC},V_{B}|\Psi\rangle_{ABC}),$ hence $\sum_i q_{i}T(U_{A}|\psi_{i}\rangle_{AB},V_{B}|\psi_{i}\rangle_{AB})\leq T(U_{A}|\Psi\rangle_{ABC},V_{B}|\Psi\rangle_{ABC}).$ This concludes the proof.

This result has a nice operational interpretation. The left-hand side $T(U_{A}\rho_{AB}U_{A}^\dagger,\rho_{AB}),$ quantifies the probability that one can tell the local unitary $U_{A}$ has been applied to $\rho_{AB}.$ The right-hand side $T(U_{A}|\Psi\rangle_{ABC},V_{B}|\Psi\rangle_{ABC}),$ quantifies the probability that someone given access to the full purified state can distinguish the application of $U_A$ from a unitary action $V_{B}$ on the other subsystem.

Despite the operational meaning, violation of this inequality does not appear to correspond directly with non-classical performance in the considered delocalised-interaction games. For pure states it does, but we can see this does not extend to mixed states with the Werner state example. For the PNP and BD games we have non-classical performance for $a>\frac{1}{2},$ and $a>\frac{1}{3}$ respectively. However, numerically we find violation of the trace distance inequality only for $a>\frac{7}{10}.$ Despite this, it is interesting to note that if we add the additional condition that A and B must keep a perfect record of what they received, or that the Bell states must not be decohered at all, then non-classical performance implies violation of the inequality for both games. These correspond to the extreme cases, where keeping a perfect record means having the left-hand side of the trace distance inequality equal to $1,$ and no decoherence requires the right-hand side to equal $0.$

To show this we first consider the PNP game subject to one of the following conditions
\begin{enumerate}
    \item A and B must always correctly record whether there was a particle and they win if C projects onto his original state.
    \item A and B must ensure that C always projects onto his original state and they win when they correctly record whether there was a particle.
\end{enumerate}
Note that these can both be considered tactical choices for the general PNP game.

Taking condition 1, A and B aim to achieve the largest possible win probability $p_{\text{pnp1}}$ subject to a perfect record condition. They must maximise
\begin{equation}\label{Eq:Con1Prob&Con}
\begin{split}
    &p_{\text{pnp1}}(\rho_{AB}) = \frac{3}{4} + \frac{1}{4}\text{Re}[\text{Tr}(U_{A}\rho_{AB}V_{B}^\dagger)], \\
    &\text{s.t. } T(\rho_{AB},\frac{1}{2}(U_{A}\rho_{AB}U_{A}^\dagger + V_{B}\rho_{AB}V_{B}^\dagger))=1.
\end{split}
\end{equation}
Note that in a finite-size Hilbert space condition 1 implies a restriction of the states $\rho_{AB}$ that can be used to play this version of the game. In addition, we know that this game has the same bound as the general game, since we showed that for pure states the best tactic was to choose unitaries that map the initial state to an orthogonal state, in other words satisfying condition 1.

We can use our trace distance inequality to derive the classical bound for the win probability. Condition 1 and convexity of the trace distance implies that $T(\rho_{AB},U_{A}\rho_{AB}U_{A}^\dagger)=1.$ If $\rho_{AB}$ is a separable state then by the trace distance inequality we have $T(U_{A}|\Psi\rangle_{ABC},V_{B}|\Psi\rangle_{ABC})=1,$ which can be written as $(1-|\text{Tr}(U_{A}\rho_{AB}V_{B}^\dagger)|^2)^{1/2}=1,$ and therefore $|\text{Tr}(U_{A}\rho_{AB}V_{B}^\dagger)|=0.$ Using this in the expression for the win probability given in Eq.~(\ref{Eq:Con1Prob&Con}) we have that for all separable states the maximum win probability is $\frac{3}{4}.$ Therefore we can conclude that non-classical performance in the game implies violation of the trace distance inequality.

We now turn to consider condition 2. For this the entanglement bound is different from the general PNP game, so we proceed to derive it for qubits as follows.
Taking condition 2 the problem is to maximise
\begin{equation}\label{Eq:Con2Prob&Con}
\begin{split}
    &p_{\text{pnp2}}(\rho_{AB}) = \frac{1}{2}[1+T(U_{A}\rho_{AB}U_{A}^\dagger, \rho_{AB})], \\
    &\text{s.t. } \text{Re}[\text{Tr}(U_{A}\rho_{AB}V_{B}^\dagger)]=1.
\end{split}
\end{equation}
Note that non-trivially satisfying the condition of Eq.~(\ref{Eq:Con2Prob&Con}) implies a restriction to states that can non-trivially satisfy  $U_{A}\rho_{AB}=V_{B}\rho_{AB},$ which were termed strongly anonymous (SA) in \cite{paige2019quantum} and shown to be the ``maximally correlated'' states studied by Rains~\cite{rains1999bound}.

We make use of the fact that the form of states that can satisfy $U_{A}\rho_{AB}=V_{B}\rho_{AB}$ was given in Ref.~\cite{paige2019quantum}. We have that a general two qubit SA state can be written in the eigenbasis of the local unitaries $U_{A},V_{B},$ as
\begin{equation*}
    \rho_{AB} = \rho_{00}|00\rangle\langle 00| + \rho_{01}|00\rangle\langle 11| + \rho_{01}^* |11\rangle\langle 00| + (1-\rho_{00})|11\rangle\langle 11|.
\end{equation*}
Since this is in the eigenbasis of the local unitaries, the unitary action $U_{A}$ can only introduce off diagonal phases, so all possible $U_{A}\rho_{AB}U_{A}$ can be written as
\begin{equation*}
    U_{A}\rho_{AB}U_{A} = \rho_{00}|00\rangle\langle 00| + e^{i\phi} \rho_{01}|00\rangle\langle 11| + e^{-i\phi} \rho_{01}^* |11\rangle\langle 00| + (1-\rho_{00})|11\rangle\langle 11|. 
\end{equation*}
The trace distance $T(U_{A}\rho_{AB}U_{A},\rho_{AB})$ is then calculated via the standard formula $T(\rho,\sigma)=\frac{1}{2}\text{Tr}(\sqrt{(\rho-\sigma)^\dagger (\rho-\sigma)})$. Using an appropriate basis we can write the states in matrix form and have
\begin{equation*}
    U_{A}\rho_{AB}U_{A} - \rho_{AB} = 
\begin{bmatrix}
    0	   & \rho_{01}(e^{i\phi}-1) \\
    \rho_{01}^*(e^{-i\phi}-1)    & 0
\end{bmatrix}.
\end{equation*}
from which it follows that
\begin{equation*}
    T(U_{A}\rho_{AB}U_{A},\rho_{AB}) = \sqrt{2(1-\cos\phi)}|\rho_{01}|.
\end{equation*}
As expected this is clearly maximised when $\cos\phi=-1$ and obtains a maximal value of $2|\rho_{01}|.$

The concurrence can be calculated as follows. Using $\tilde{\rho} = (Y \otimes Y) \rho^{*} (Y \otimes Y),$ we have
\begin{equation*}
    \tilde{\rho}_{AB} = (1-\rho_{00})|00\rangle\langle 00| + \rho_{01}|00\rangle\langle 11| + \rho_{01}^* |11\rangle\langle 00| + \rho_{00}|11\rangle\langle 11|.
\end{equation*}
In matrix form we find
\begin{equation*}
    \rho_{AB} \tilde{\rho}_{AB} = 
\begin{bmatrix}
    \rho_{00}(1-\rho_{00})+|\rho_{01}|^2	   & 2\rho_{00}\rho_{01} \\
    2(1-\rho_{00})\rho_{01}^*    & \rho_{00}(1-\rho_{00}) + |\rho_{01}|^2
\end{bmatrix}.
\end{equation*}
The eigenvalues are given by
$\lambda_{\pm} = \rho_{00}(1-\rho_{00}) + |\rho_{01}|^2 \pm 2\sqrt{\rho_{00}(1-\rho_{00})|\rho_{00}|^2},$
and the square roots of these are $\sqrt{\lambda_{\pm} } = \sqrt{\rho_{00}(1-\rho_{00})} \pm |\rho_{01}|.$ The fact that density matrices are positive semi-definite means that $\sqrt{\rho_{00}(1-\rho_{00})} \geq |\rho_{01}|$ and therefore $\sqrt{\lambda_{+}}\geq\sqrt{\lambda_{-}},$ so the concurrence is calculated as
\begin{equation*}
    \sqrt{\lambda_{+}}-\sqrt{\lambda_{-}} = 2|\rho_{01}|.
\end{equation*}
But this is precisely the maximum value of $T(U_{A}\rho_{AB}U_{A},\rho),$ therefore we have shown that for qubit states that can be used to play this game we have maximum win probability
\begin{equation*}
    p_{\text{pnp2}}^{\text{m}}(\rho_{AB}) = \frac{1}{2}[1+C(\rho_{AB})].
\end{equation*}
This is clearly reminiscent of the BD game concurrence bound, but note that not all Bell diagonal states are SA states, and therefore some states that saturate the BD game concurrence bound cannot even be used to play the PNP game with condition 2.

We could consider higher dimensional cases. The form for SA states (with no degeneracy) is  $\sum_{i,j}\rho_{i,j}|ii\rangle\langle jj|,$ and our local unitaries are of the form $U_{A}=\sum_{k}e^{i\phi_{k}}|k\rangle_{A}\langle k|,$ $V_{B}=\sum_{k}e^{i\phi_{k}}|k\rangle_{B}\langle k|.$ For the two qubit case we had two unitary phases to vary one relative phase in the state (the off-diagonal term). This made the maximization simple as we just arranged the relative phase to be $\pi.$ However, we straight away can see that for high $d$ dimensions we shall clearly run into a problem with the number of unitary phases scaling as only $d,$ whereas the number of relative phases scales as $(d^2-d)/2.$ Furthermore, even in the case $d=3,$ where the scaling is not a problem since $(3^2-3)/2=3,$ the result from the optimization appears to be a long and unenlightening expression.

Now that we know the classical bound it is straightforward to demonstrate that non-classical performance implies violation of the trace inequality. Condition 2 implies that $T(U_{A}|\Psi\rangle,V_{B}|\Psi\rangle)=0,$ and hence for all separable states under this condition we have $T(U_{A}\rho_{AB}U_{A}^\dagger, \rho_{AB})=0,$ giving a maximum win probability of $\frac{1}{2},$ i.e. without some form of entanglement all A and B can do is guess. Therefore we have the claimed result.

What of the BD game. For condition 1 the problem becomes the maximisation of
\begin{equation}\label{Eq:BDCon1Prob&Con}
\begin{split}
    &p_{\text{bd1}}(\rho_{AB}) = \frac{1}{2} + \frac{1}{4}\text{Re}[\text{Tr}(U_{A}\rho_{AB}V_{B}^\dagger+U_{A}\rho_{AB}V_{B})], \\
    &\text{s.t. } T(\frac{1}{2}(\rho_{AB} + U_{A}V_{B}\rho_{AB}U_{A}^\dagger V_{B}^\dagger) ,\frac{1}{2}(U_{A}\rho_{AB}U_{A}^\dagger + V_{B}\rho_{AB}V_{B}^\dagger))=1.
\end{split}
\end{equation}
The classical bound for this game is the guessing probability of $\frac{1}{2},$ as we would expect. Applying the convexity of the trace distance to the condition of Eq.~(\ref{Eq:BDCon1Prob&Con}) leads through to $T(\rho_{AB},U_{A}\rho_{AB}U_{A}^\dagger)=1.$ Then by applying the same arguments as above with the trace distance inequality we conclude that $|\text{Tr}(U_{A}\rho_{AB}V_{B}^\dagger)|=0$ and $|\text{Tr}(U_{A}\rho_{AB}V_{B})|=0,$ therefore the maximum win probability is $\frac{1}{2}.$

For the BD game under condition 2 we have to maximise
\begin{equation}\label{Eq:BDCon2Prob&Con}
\begin{split}
    &p_{\text{bd2}}(\rho_{AB}) = \frac{1}{2}[1+T(U_{A}\rho_{AB}U_{A}^\dagger, \rho_{AB})], \\
    &\text{s.t. } \text{Re}[\text{Tr}(U_{A}\rho_{AB}V_{B}^\dagger)]= \text{Re}[\text{Tr}(U_{A}\rho_{AB}V_{B})]=1.
\end{split}
\end{equation}
Unlike in the PNP game, this time there is no need to derive a new bound, the guessing probability of $\frac{1}{2}$ is still the best we can do classically in accordance with Eq.~(\ref{Eq:BDCon2Prob&Con}) (since it is the do nothing strategy). By precisely the same argument as before we then see that the trace distance inequality implies a maximum win probability of $\frac{1}{2}.$ Thus we have found that in all four cases, non-classical performance implies violation of the trace distance inequality.

\end{document}